\documentclass[aip, pop, amsmath,amssymb, reprint]{revtex4-2}

\usepackage{amsmath,amssymb,mathtools}
\usepackage{bm}
\usepackage{booktabs}
\usepackage{array}
\usepackage{multirow}
\usepackage{dcolumn}
\usepackage{graphicx}
\usepackage{hyperref}
\usepackage{color}
\usepackage{physics}
\usepackage{enumitem}
\usepackage{braket}
\newcolumntype{d}[1]{D{.}{.}{#1}}

\begin{document}

\title{Bound states of the hydrogen-like atomic systems in plasma environments}

\author{Fatma Zohra Khaled}
\email{fatmazohra.Khaled@univ-batna.dz}
\affiliation{LPRIM, Department of Physics, University of Batna I, 05000, Algeria}

\author{Mustapha Moumni}
\email{m.moumni@univ-batna.dz}
\affiliation{LPRIM, Department of Physics, University of Batna I, 05000, Algeria}
\affiliation{LPPNNM, Department of Matter Sciences, University of Biskra, 07000, Algeria}

\author{Mokhtar Falek}
\email{falek.mokhtar@univ-khenchela.dz}
\affiliation{LPPNNM, Department of Matter Sciences, University of Biskra, 07000, Algeria}
\affiliation{Faculty of Technology, University of Khenchela, 40000, Algeria}

\date{\today}

\begin{abstract}
We conduct a non-relativistic study of plasma screening effects on hydrogen-like atomic systems using  the Screened Coulomb Potential (SCP i.e. Yukawa potential). The radial Schrödinger equation is first reduced to a bi-confluent Heun (BCH) equation for the Killingbeck potential which is the truncated version of the SCP, and we write the exact BCH eigenfunctions and eigenenergies. We then study the limitations of these BCH solutions and obtain an analytic description valid for weak to moderate screening using the BCH functional form. The corrected eigenfunctions are constructed order by order up to $\mathcal{O}(k^3)$; they are expressed in terms of the Laguerre polynomials and reduces exactly to Coulomb eigenfunctions when $k=0$. Using these Functions, the energy spectrum is computed via two analytic methods: (i) direct evaluation of the full Yukawa Hamiltonian expectation value, and (ii) the Hellmann-Feynman theorem, yielding integral representation. All methods presented here provide explicit analytical formulas for both wavefunctions and eigenenergies valid for different ranges of the screening. These methods establish a powerful analytic framework for studying confined quantum systems. The thermodynamic properties are also derived using the BCH formulation.
\end{abstract}

\maketitle

\section{Introduction}
\label{sec:introduction}

The study of quantum confinement in atomic systems in charged environments like plasma, is considered as one of the fundamental topics in physics\cite{Martinez2018}. Over past decades, it has attracted considerable interest, with wide-ranging applications astrophysical environments\cite{Leckrone1993,Kuramitsu2012}, semiconductors and quantum dots\cite{Kwon2006,Genkin2010}, material processing \cite{Penkov2015}, fusion processes\cite{Nishikawa2000}. Among the various models proposed to describe the interaction between charged particles and plasma, the exponential screened Coulomb potential (SCP)\cite{Messina2003,Sil2009}, also known as the Debye-Hückel potential \cite{Debye1923} or Yukawa potential\cite{Yukawa1935}. This potential reads $V(r)=-(Ze^{2}/r)e^{-kr}$ where $k$ is the screening parameter and it is defined as the inverse of the Debye wavelength $\lambda_{D}$. We find also some generalisation of this SCP, like the exponential cosine screened Coulomb potential \cite{Shukla2012,Soylu2012,Qi2017}; the generalized exponential screened Coulomb potential \cite{Ikhdair2007}; and
the more generalized exponential screened Coulomb potential \cite{Soylu2012,Sever1987}. More recently, the Coulomb potential at finite temperature has also been investigated \cite{Zhao2017}.

Despite its physical inportance, the Schrödinger equation for the SCP cannot be solved exactly in terms of elementary functions. Consequently, a variety of methods have been employed: Ritz variational technique\cite{Paul2009}, the finite difference method\cite{Rogers1970}, the generalized pseudospectral (GPS) method\cite{Roy2013}, the asymptotic iteration method (AIM)\cite{Gonul2006}, the Nikiforov-Uvarov method\cite{Hamzavi2012}, supersymmetric quantum mechanics (SUSY) \cite{Lee2000,Napsuciale2021}, Numerical\cite{Nasser2011}, hypervirial Pad\'{e} approximation method\cite{Hirschfelder1960,Killingbeck1978,Grant1979,Lai1982}, and the 1/N expansion method\cite{Lai1982}. While these approaches provide valuable numerical results, they often lack explicit analytic expressions for the wavefunctions and energies as functions of the screening parameter.

A common simplification is to expand the exponential factor according to parameter $k$. Here we do this expansion up to the 3-order and get the cubic truncated potential of the Killingbeck type
\begin{equation*}
V\left( r\right) =Ze^{2}\left( -\frac{1}{r}+k-\frac{k^{2}r}{2}+\frac{k^{3} r^{2}}{6}\right)+\mathcal{O}(k^{4}) 
\end{equation*}
For this potential, the radial Schrödinger equation can be transformed into a bi-confluent Heun (BCH) equation. We write the exact solutions of the BCH equation for all states corresponding to the radial quantum number $n_r =0$ and $n_r =1$, providing closed-form energy expressions for the Yukawa system. However these exact solutions suffer from a drawback: they do not reduce to the Coulomb eigenfunctions when $k\to 0$ despite the fact that they are analytic solutions to the Killingbeck potential which can be considered as a generalization of the Coulomb one.

To overcome this limitation while preserving the analytic character of the solution, we propose an alternative analytic approach that retains the \textit{functional form} of the BCH ansatz but without their rigid truncation relations. Instead, we treat the potential parameters and the polynomial coefficients as analytic functions of $k$ and determine them order by order by requiring the Schrödinger equation to be satisfied and the Coulomb limit to be recovered. This yields a set of corrected eigenfunctions that are analytic in $k$ and reduce exactly to the Coulomb solutions when $k=0$; the construction is systematic and can be extended to any order. using these corrected BCH wavefunctions, we compute the Yukawa energy spectrum via two additional independent analytic methods:
\begin{enumerate}[label=(\roman*)]
        \item direct expectation value of the full Yukawa Hamiltonian, which give closed-form expressions in terms of hypergeometric functions
        \item the Hellmann-Feynman theorem, which yields an integral representation of the solutions.
\end{enumerate}
Both methods give the same perturbative expansion, which agree with the standard Rayleigh-Schrödinger perturbation theory up to $\mathcal{O}(k^{3})$. The resulting formulas are explicit analytic functions of $k$ and allow immediate numerical evaluations.

Compared to purely numerical or variational methods, our approach offers several advantages:
\begin{enumerate}[label=(\roman*)]
        \item explicit analytic expressions for wavefunctions and energies ou to the desired order.
        \item systematic improvement by including higher order terms
        \item Direct calculation of expectation values and transition probabilities
        \item smooth Coulomb limit by construction
        \item closed-form Laguerre representations of the wavefunctions that can be used for further analytic works and also for variational computations with one and two parameters.
\end{enumerate}

In addition, the thermodynamic properties of the SCP are also studied in this article. The analytical expression for the partition function and the corresponding thermodynamic quantities are derived using the Euler-Maclaurin summation formula. These include free energy, mean energy, entropy, and heat capacity. Previously, different authors have investigated thermodynamic properties for several physical systems\cite{Okorie2018,Ikot2019,Okorie2020,Inyang2021}. 

The paper is organized as follows. Section~\ref{sec:formalism} presents the theoretical formalism including the exact BCH solutions and the construction of corrected BCH wavefunctions with the corresponding analytic methods. Section~\ref{sec:properties} discussses the thermodynamic properties derived from the partition function. Section~\ref{sec:results} contains the results and discussions, including numerical tables and graphical illustrations. Finally, Section~\ref{sec:conclusion} concludes the paper. Technical details are provided in the appendices.
\section{Theoretical Formalism}
\label{sec:formalism}

The non-relativistic Hamiltonian of a hydrogen-like atomic system immersed in a plasma medium described by the screened Coulomb potential $\left(SCP\right) $ is 
\begin{equation}
H=-\frac{\hbar ^{2}}{2\mu }\nabla ^{2}-\frac{Ze^{2}}{r}e^{-kr}
\label{eq:hamiltonian}
\end{equation}
where $Z$ is the atomic number, $e^{2}=\frac{q^{2}}{4\pi \varepsilon _{0}}$, $\mu $ is the reduced mass and $k=1/\lambda$ is the screening parameter.

In spherical coordinates, the time-independent Schrödinger equation becomes 
\begin{equation}
\left( -\frac{\hbar ^{2}}{2\mu }\nabla ^{2}-\frac{Ze^{2}}{r}e^{-kr}\right)\psi \left( r,\theta ,\varphi \right) =E\psi \left( r,\theta ,\varphi
\right) \,
\label{eq:radial_schrodinger}
\end{equation}
Using the separation $\psi \left( r,\theta,\varphi \right) =\frac{U_{n_{r},\ell}\left( r\right) }{r}Y_{\ell}^{m_{\ell}}\left( \theta,\varphi \right)$, the angular part gives the usual spherical harmonics,
\begin{equation}
L^{2}Y_{\ell}^{m_{\ell}}\left( \theta ,\varphi \right) =\hbar ^{2}\ell\left(\ell+1\right) Y_{\ell}^{m_{\ell}}\left( \theta ,\varphi \right) 
\end{equation}
and the radial equation reads
\begin{equation}
\left( \frac{d^{2}}{dr^{2}}-\frac{l\left( l+1\right) }{r^{2}}+\frac{2\mu
Ze^{2}}{\hbar ^{2}r}e^{-kr}+\frac{2\mu E_{n_{r},\ell}}{\hbar^{2}}\right)
U_{n_{r},\ell}\left( r\right) =0\,
\end{equation}
We focus on small enough values of the screening parameter $k$ (i.e., large Debye length $\lambda_D$), where the Taylor expansion up to third order is accurate
\begin{equation}
V\left( r\right) =Ze^{2}\left( -\frac{1}{r}+\frac{1}{\lambda }-\frac{r}{2\lambda ^{2}}+\frac{r^{2}}{6\lambda ^{3}}\right) 
\end{equation}
This gives the cubic-truncated potential (Killingbeck type). Rearranging the Schrödinger equation gives 
\begin{equation}
\left[\frac{d^2}{dr^2}+C_0+C_1r+C_2r^2+\frac{C_3}{r}+\frac{C_4}{r^2}\right]U_{n_{r},\ell}(r)=0,
\label{eq:radial_c_coeffs}
\end{equation}
where
\begin{align}
C_{0}&=\frac{2\mu }{\hbar ^{2}}\left( E_{n,\ell}-\frac{Ze^{2}}{\lambda }\right),C_{1}=\frac{\mu }{\hbar ^{2}}\frac{Ze^{2}}{\lambda ^{2}},C_{2}=-\frac{\mu }{3\hbar ^{2}}\frac{Ze^{2}}{\lambda ^{3}}, \notag \nonumber \\ 
C_{3}&=\frac{2\mu Ze^{2}}{\hbar ^{2}}\text{ and }C_{4}=-\ell\left( \ell+1\right) 
\end{align}

\subsection{Exact Biconfluent Heun Solutions}
To solve Eq.~\ref{eq:radial_c_coeffs}, we use the ansatz
\begin{equation}
U_{n_{r},\ell}(r)=r^{\alpha }e^{-\left( \beta r+\gamma r^{2}\right) }g_{n_{r},\ell}(r)
 \label{eq:exact_killingbeck}
\end{equation}
where $\alpha ,\beta $ and $\gamma $ are defined by
\begin{equation}
\text{ }\alpha =\ell+1,\text{ }\beta =\frac{-C_{1}}{2\sqrt{-C_{2}}}\text{ and }
\gamma =\frac{\sqrt{-C_{2}}}{2}
\end{equation}
Substituting Eq.~\ref{eq:exact_killingbeck} in Eq.~\ref{eq:radial_c_coeffs} leads to
\begin{align}
&g_{n_{r},\ell}''(r)+\left(\frac{2\alpha}{r}-2\beta-4\gamma r\right)g_{n_{r},\ell}'(r) \label{eq:g_equation_r} \\
&+\left[C_0+\beta^2-2\gamma(2\alpha+1)+\frac{C_3-2\alpha\beta}{r}\right]g_{n_{r},\ell}(r)=0. \nonumber
\end{align}
Introducing the dimensionless variable $\rho =\sqrt[4]{-C_{2}}r$ transforms \ref{eq:g_equation_r} into the biconfluent Heun equation\cite{Ronveaux1995}
\begin{align}
\rho{g}_{n_{r},\ell}''(\rho)&+\left(1+\alpha'-\beta'\rho-2\rho^2\right){g}_{n_{r},\ell}'(\rho) \label{eq:BCH}\\
&+\left[(\gamma'-\alpha'-2)\rho-\frac{\delta'+(1+\alpha')\beta'}{2}\right]{g}_{n_{r},\ell}(\rho)=0, \nonumber
\end{align}
where the parameters are given by
\begin{align}
\alpha'&=2\ell+1, \gamma'=\frac{\beta^2+C_0}{(-C_2)^{1/2}},\nonumber\\
\beta'&=\frac{2\beta}{\sqrt[4]{-C_2}}\text{ and }\delta'=\frac{-2C_3}{\sqrt[4]{-C_2}}.
\label{eq:BCH_coefficients}
\end{align}
The regular solution of Eq.~(\ref{eq:BCH}) around the origin ($\rho =0$) can be expressed as a power series \cite{Ronveaux1995,Khaled2025}
\begin{align}
g_{n_{r},\ell}(\rho)&=H_{b}\left( \alpha ^{\backprime },\beta^{\backprime },\gamma \backprime ,\delta ^{\backprime },\rho \right)\nonumber\\
&=\sum_{n_{r}=0}^{\infty}a_{n_{r}}\frac{\Gamma(1+\alpha')}{\Gamma(1+\alpha'+n_{r})}\frac{\rho^{n_{r}}}{n_{r}!},
\label{eq:g_series}
\end{align}
with $a_{0}=1$ and  $a_{1}=\frac{1}{2}\left( \delta ^{\backprime }+\beta ^{\backprime }\left( 1+\alpha
^{\backprime }\right) \right) $. The remaining coefficients satisfy the recurrence relation
\begin{align}
a_{n_{r}+2}=&\left[\frac{1}{2}\bigl(\delta'+\beta'(1+\alpha')\bigr)+\beta'(n_{r}+1)\right]a_{n_{r}+1} \nonumber\\
&-(n_{r}+1)(n_{r}+1+\alpha')(\Delta-2n_{r})\,a_{n_{r}},
\label{eq:recurrence}
\end{align}
where $\Delta =$ $\gamma \backprime -\alpha ^{\backprime }-2$.

The series truncates to a polynomial of degree $n_{r}$ (i.e., the bound-state condition) when both following conditions are satisfied\cite{Khaled2025}
\begin{align}
\Delta &= 2n_{r},\qquad n_{r}=0,1,2,\ldots=n-\ell-1,
\label{eq:trunc1}\\
a_{n_{r}+1} &= 0.
\label{eq:trunc2}
\end{align}
From the first condition Eq.~(\ref{eq:trunc1}), we obtain the energies
\begin{equation}
E_{n_{r},\ell}=\frac{\hbar ^{2}}{\mu }\left[ \sqrt{-C_{2}}\left( n_r+\ell+\frac{3}{2}\right) -\frac{1}{2}\left( \frac{C_{1}}{2\sqrt{-C_{2}}}\right) ^{2}\right] +\frac{Ze^{2}}{\lambda }
\label{eq:energy}
\end{equation}
The corresponding wave function is 
\begin{align}
& \psi \left( r,\theta ,\varphi \right) =A_{n,\ell}r^{l}e^{\frac{1}{2}\sqrt{\frac{3\mu }{\hbar ^{2}}\frac{Ze^{2}}{\lambda }}\left( r-\frac{r^{2}}{3\lambda }
\right) } \nonumber\\
&\times H_{b}\left( \alpha ^{\backprime },\beta ^{\backprime },\gamma \backprime,\delta ^{\backprime },\sqrt[4]{-C_{2}}r\right) Y_{\ell}^{m_{\ell}}\left( \theta,\varphi \right) ,
\label{eq:wave function}
\end{align}
where $A_{n,\ell}$ is the normalization constant.

The second truncation condition Eq.~(\ref{eq:trunc2}) imposes additional constraints. In what follows, we analyze this condition for the first two radial quantum numbers $n_r$.
\subsection{For the case $\left( n_{r}=0\right)$}
\label{subsec:case 1}
For $n_{r}=0$, we have $\left( \Delta =0\right)$; the condition $a_{1}=0$ gives
\begin{equation}
\frac{C_{1}}{2\sqrt{-C_{2}}}=\frac{-C_{3}}{2\left( \ell+1\right) }=\frac{\mu Ze^{2}}{\hbar ^{2}\left( \ell+1\right) }
\label{eq:recurrence1}
\end{equation}
This links the three parameters $C_{1}$, $C_{2}$, and $C_{3}$. Using (\ref{eq:recurrence1}  Eqs.~(\ref{eq:energy} yields the energy eigenvalues for the states $(n_{r}=0,\ell)$\cite{Khaled2025}
\begin{equation}
E_{0,\ell}=-\frac{\mu \left( Ze^{2}\right) ^{2}}{2\hbar ^{2}\left(\ell+1\right)
^{2}}+\frac{Ze^{2}}{\lambda }\left( 1+\sqrt{\frac{\hbar ^{2}}{3\mu Ze^{2}\lambda }}\left( \ell+\frac{3}{2}\right) \right) 
\label{eq:energy1}
\end{equation}
and the radial wavefunctions simplify to 
\begin{equation}
R_{0,\ell}(r)=A_{0,\ell}r^{\ell}e^{\frac{1}{2}\sqrt{\frac{3\mu }{\hbar ^{2}}\frac{Ze^{2}}{\lambda }}\left( r-\frac{r^{2}}{3\lambda }\right) }
\label{eq:wave function1}
\end{equation}
with $A_{0,\ell}$ given in  Appendix~(\ref{app:Norms-BCH}).
\subsection{For the case $\left(n_{r}=1\right)$}
\label{subsec:case 2}
For $n_{r}=1$ $\left( \Delta =2\right) $, the condition $a_{1}=0$ leads to\cite{Khaled2025}
\begin{align}
a_{2}=&\frac{1}{2}\left( \beta'+\frac{1}{2}\left( \delta '+\beta'\left( 1+\alpha'\right)
\right) \right) \left( \delta'+\beta'\left(1+\alpha'\right) \right) \nonumber\\
& -2\left( \alpha'+1\right) =0 ,
\label{eq:recurrence2}
\end{align}
Therefore, we get the constraint
\begin{widetext}
\begin{equation}
\frac{C_{1}}{2\sqrt{-C_{2}}}=\frac{-C_{3}}{4\left( \ell+1\right) \left(
\ell+2\right) } \left( \left( 2\ell+3\right) \pm \sqrt{16\left( \ell+1\right)^{2}\left(\ell+2\right) \frac{\sqrt{-C_{2}}}{C_{3}^{2}}+1}\right) ,
\label{eq:constraint2}
\end{equation}
\end{widetext}
Substituting the appropriate branch (the one that recovers the correct Coulomb limit) gives\cite{Khaled2025}
\begin{widetext}
\begin{equation}
E_{1,\ell}=-\frac{\mu \left( Ze^{2}\right) ^{2}}{8\hbar ^{2}(\ell+1)^{2}(\ell+2)^{2}}
\left( 2\ell+3-\sqrt{12\left( \frac{\hbar ^{2}}{3\mu Ze^{2}\lambda }\right)^{3/2}(\ell+1)^{2}(\ell+2)+1}\right) ^{2}+\frac{Ze^{2}}{\lambda }\left( 1+\sqrt{\frac{\hbar ^{2}}{3\mu Ze^{2}\lambda }}\left( \ell+\frac{5}{2}\right) \right),
\label{eq:energy2}
\end{equation}
\end{widetext}
The corresponding radial wavefunction is
\begin{align}
R_{_{1,\ell}}\left( r\right) =&A_{1,\ell}r^{\ell}e^{\frac{1}{2}\sqrt{\frac{3\mu }{\hbar ^{2}}\frac{Ze^{2}}{\lambda }}\left( r-\frac{r^{2}}{3\lambda }\right)}\nonumber\\
&\times\left( 1+\frac{\Gamma \left( 2\ell+2\right) }{\Gamma \left( 2\ell+3\right) }a_{1}\left( \sqrt[4]{-C_{2}}r\right) \right).
\label{eq:wave function2}
\end{align}
The normalization constant $A_{1,\ell}$ is given in Appendix~(\ref{app:Norms-BCH}.

For $n_{r}=2$, the constraint $a_3 =0$ is more complicated; the explicit expression is out of the present discussion.

We can observe from the expressions obtained for the system's energy that, for each value of $n_{r}$, a unique energy is obtained under its specific constraint, differing from those associated with other values of $n_{r}$. The  $\ell$-dependence of the energy is thus encoded in the physical mapping $n=n_{r}+\ell+1$: for a given $n$, states with larger $\ell$ correspond to smaller $n_{r}$, and hence to different Heun polynomials. The second truncation condition, Eq.~(\ref{eq:trunc2}), imposes an additional constraint that explicitly introduces $\ell$ through  $\alpha'=2\ell+1$ and must therefore be treated separately for each  $(n, \ell) $ pair. Eq.~((\ref{eq:energy}) does not represent the exact Yukawa energy; rather, it corresponds to the energy condition associated with the third-order polynomial approximation of the Yukawa potential (Killingbeck-type).

The Biconfluent Heun solutions represent the most complete analytical treatment that can be derived for the truncated cubic SCP. However, despite being exact analytical solutions of the Killingbeck potential, they remain approximate solutions for the SCP potential, as demonstrated by the numerical results reported in Tables~\ref{tab:comp_energy_n0},\ref{tab:comp_energy_n1}.

A more fundamental limitation accuracy of the exact BCH polynomial treatment emerges from the behavior of associated special functions in the Coulomb limit. Specifically, the Biconfluent Heun equation Eq.~(\ref{eq:BCH}) exhibits a well-known degenerate limit when both  $\beta'=0$, $\delta'=0$. In that limit where $C_1=C_3=0$, it reduces to the confluent hypergeometric equation\cite{Ronveaux1995}
\begin{equation}
H_B(\alpha',0,\gamma',0,\rho) = {}_1F_1\left( \frac{1}{2}+\frac{\alpha'-\gamma'}{4},1+\frac{\alpha'}{2}; \rho^2 \right)
\label{eq:BCH_kummer_limit}
\end{equation}
whose polynomial solutions are expressed in terms of the generalized Laguerre polynomials $L_{n_r}^{\alpha'/2}(\rho^2)$, i.e., the radial eigenfunctions of the harmonic oscillator.

When $k\to 0$, the SCP reduces to the pure Coulomb potential $-1/r$ and both $C_1$ and $C_2$ vanish as well as $\beta'$ and $\gamma'$; the BCH equation cannot reduce to the confluent hypergeometric equation of the Coulomb case. Consequently, the exact BCH polynomial truncation condition does not allow a smooth transition to the Coulomb eigenfunctions - a serious drawback for any perturbative or semi-classical treatment of weak screening.

To overcome this difficulty, we develop an alternative approach that retains the \textit{functional form} of the BCH solutions in \ref{eq:exact_killingbeck}, but abandons the rigid BCH truncation conditions \ref{eq:trunc1} and \ref{eq:trunc2}. Instead, we treat the parameters $\alpha$, $\beta$ and the coefficients $a_{j}$ in the polynomial form of $g(r)$ as unknown functions of the screening parameter $k$, to be determined by the physical requirements that the wavefunctions satisfies the Schrödinger equation order by order in $k$, and that the energy eigenvalues coincide with those obtained from a stable perturbative expansion around the Coulomb problem.

Because the Hamiltonian depends analytically on $k$, we expand all quantities in integer power of $k$ (no half-integer as in the exact BCH solutions) and enforce the Schrödinger equation recursively. This yields \textit{analytic} eigenfunctions that reduce exactly to the Coulomb ones when $k$ vanishes and reproduce the correct energy spectrum up to $\mathcal{O}(k^{3})$.

Thus, while the exact BCH polynomial solutions are limited to the strongly screened regime and cannot recover the Coulomb limit, the alternative approach described above provide a consistent, analytic treatment valid for all screening strengths.
\subsection{Inspired BCH Eigenfunctions}
\label{subsec:perturbative}
The exact BCH polynomial solutions discussed above have a fundamental shortcoming: they do not reduce to the correct Coulomb eigenfunctions when the screening parameter $k \rightarrow 0$; this limitation is rooted in the algebraic structure of the Heun equation. To overcome this, keep the \textit{functional form} of the BCH ansatz  \ref{eq:exact_killingbeck}, we relax the rigid truncation conditions (\ref{eq:trunc1}-\ref{eq:trunc2}) and we write the parameters $\alpha$, $\beta$ and the coefficients $a_{j}$ as analytic functions of $k$. Because the Hamiltonian depends polynomially on $k$, the physical eigenfunctions are analytic in $k$; hence this expansion is well defined. The fundamental requirement is that the energies obtained via these eigenfunctions coincide with those coming from the usual perturbative approach of the Yukawa potential around the Coulomb problem up to the $3-$order in $k$.
\begin{align}
E_{n_r,\ell}=& E^{(0)}_{n_r,\ell} + \langle \psi^{(0)}_{n_r,\ell} | Ze^{2} | \psi^{(0)}_{n_r,\ell} \rangle - k^2 \langle \psi^{(0)}_{n_r,\ell} | \frac{ Ze^{2}}{2}r | \psi^{(0)}_{n_r,\ell} \rangle \nonumber \\
+& k^3 \langle \psi^{(0)}_{n_r,\ell} | \frac{ Ze^{2}}{6}r^2 | \psi^{(0)}_{n_r,\ell} \rangle + \mathcal{O}(k^4),
\end{align}
where $\psi^{(0)}_{n_r,\ell}(r)$ and $E^{(0)}_{n_r,\ell}$ are the Coulomb solutions:
\begin{equation}
\psi^{(0)}_{n_r,\ell}(r) = \mathcal{N}^{(0)}_{n_r,\ell}r^{l} e^{-r/(a_0(n_r + \ell + 1))} L_{n_r}^{2\ell+1}(x)
\label{eq:unter-coloumb}
\end{equation}
\begin{equation}
E^{(0)}_{n_r,\ell} = -\frac{\mu (Ze^{2})^2}{2\hbar^2(n_r + \ell + 1)^2}, \quad n = n_r + \ell + 1.
\end{equation}
and $\mathcal{N}^{(0)}_{n_r,\ell}$ is the normalization constant of the $\psi^{(0)}_{n_r,\ell}(r)$:
\begin{equation}
\mathcal{N}^{(0)}_{n_r,\ell}=\sqrt{\frac{2}{a_0(n_r+\ell+1)^2}\frac{n_r!}{(n_r+2\ell+1)!}}
\label{eq:Norm_0}
\end{equation}
Using the expressions of both $\langle r \rangle^{(0)}$ and $\langle r^2 \rangle^{(0)}$, we get the total energy to order $k^3$ (here $n=n_r+\ell+1$):
\begin{align}
E_{n_r,\ell}=&E^{(0)}_{n_r,\ell} + k E^{(1)}_{n_r,\ell} +k^2 E^{(2)}_{n_r,l} + k^3 E^{(3)}_{n_r,\ell} + \mathcal{O}(k^4) \nonumber\\
=&-\frac{ \mu (Ze^{2})^2}{2\hbar^2 n^2}+Ze^{2}k-\frac{\hbar^2}{4 \mu} \left[3 n^2 - \ell(\ell+1) \right]k^2 \nonumber\\
+&\frac{\hbar^4 n^2}{12 \mu^2Ze^{2}} \left[5 n^2 + 1 - 3\l(\ell+1)\right]k^3+ \mathcal{O}(k^4)
\label{eq:Coulomb_Energy_3}
\end{align}

The construction (detailed in Appendix \ref{app:Inspired}) yields the following corrected wavefunctions up to $\mathcal{O}(k^4)$:
\begin{align}
U_{n_r,\ell}(r)=&\mathcal N_{n_r,\ell}r^{\ell+1}e^{-r/a_0 n}\nonumber\\
\times& \left[L_{n_r}^{2\ell+1}(x)+k^2 P^{(2)}_{n_r,\ell}(x)+k^3 P^{(3)}_{n_r,\ell}(x) \right].
\label{eq:pert-scp}
\end{align}
where $a_0 = \frac{\hbar^2}{\mu Z e^2}, n = n_r + \ell + 1, x = \frac{2r}{a_0 n}$ and the polynomials $P^{(2)}_{n_r,\ell}(x)$ and $P^{(3)}_{n_r,\ell}(x)$ are given by:
\begin{align}
P_{n_r,\ell}^{(2)}(x) &= \frac{1}{4}\bigl[3n^2 - \ell(\ell+1)\bigr] L_{n_r}^{2\ell+1}(x) - \frac{n x}{2} L_{n_r-1}^{2l+2}(x), \nonumber\\[6pt]
P_{n_r,\ell}^{(3)}(x) &= \frac{n^2}{12}\bigl[5n^2 + 1 - 3\ell(\ell+1)\bigr] x^2 L_{n_r}^{2\ell+1}(x) \nonumber \\
&\quad - \frac{n}{6}\bigl[3n^2 - \ell(\ell+1)\bigr] x^2 L_{n_r-1}^{2\ell+2}(x) \nonumber \\
&\quad + \frac{n^2}{12} x^2 L_{n_r-2}^{2\ell+3}(x).
\label{eq:polynom_2-3}
\end{align}
The full expression of $N_{n_r,\ell}r$ is detailed in Appendix \ref{app:Norms-BCH}. These BCH inspired eigenfunctions satisfy the Schrödinger equation for the cubic-truncated Yukawa potential up to $\mathcal{O}(k^4)$ and reduce exactly to the Coulomb eigenfunctions when $k=0$.

We can check these solutions by computing the expectation value of the Yukawa Hamiltonian:
\begin{equation}
E_{n_r,\ell}(k) = \frac{\langle U_{n_r,\ell} | H | U_{n_r,\ell} \rangle}{\langle U_{n_r,\ell} | U_{n_r,\ell} \rangle}
\label{eq:Exp_Value_3}
\end{equation}
Substituting the expression of $U_{n_r,\ell}$ from \ref{eq:pert-scp} and expanding the numerator and denominator in powers of $k$ (using the orthogonality and recursion relations of Laguerre polynomials), we obtain after a straightforward calculation the same expression in \ref{eq:Coulomb_Energy_3} (Appendix \ref{app:Inspired}).
\subsection{Energy via Direct Expectation Values}
\label{subsec:case 2}
The energy of a stationary state is the expectation value of the Hamiltonian; So to get a better estimate of the Yukawa case energies, we will use the expression of $U_{n_r,\ell}$ from \ref{eq:pert-scp} to compute the expectation value of the full Yukawa Hamiltonian (not the truncated expansion), so we compute:
\begin{equation}
E_{n_r,\ell} = 
\frac{\displaystyle\int_0^\infty U_{n_r,\ell}(r)\left(H(Yukawa)\right)U_{n_r,\ell}(r)\,dr}
{\displaystyle\int_0^\infty |U_{n_r,\ell}(r)|^2\,dr}
\label{eq:Exp_Value_Full}
\end{equation}
where
\begin{equation}
H(Yukawa)=-\frac{\hbar^2}{2\mu}\frac{d^2}{dr^2}+\frac{\ell(\ell+1)\hbar^2}{2\mu r^2}-\frac{Ze^2}{r}e^{-kr}
\end{equation}
The integrals can be evaluated using the properties of the generating function of Laguerre polynomials or their integral representation:
\begin{equation}
L_n^{(\alpha)}(x) = \frac{1}{n!} e^x x^{-\alpha} \frac{d^n}{dx^n}\bigl(e^{-x} x^{n+\alpha}\bigr)
\end{equation}
which can be expressed in closed form with hypergeometric functions $_{2}F_{1}$ and $_{1}F_{1}$ (Appendix \ref{app:Inspired}).
We have computed these expressions numerically and the energy eigenvalues obtained from this direct expectation value method using both $2-$order and $3-$order inspired BCH wavefunctions from \ref{eq:pert-scp} are listed in columns $E_{exp2}$ and $E_{exp3}$ in Tables~\ref{tab:comp_energy_n0},\ref{tab:comp_energy_n1}. Already at second order ($E_{exp2}$) the results are in good agreement with the reference values; the inclusion of third order corrections ($E_{exp3}$) reduces the relative errors to below $0.01\%$ for most states with $\lambda_{D}\gtrsim20$. A full discussion of the convergence and comparison with other methods is deferred to later.
\subsection{Energy levels  via Hellmann-Feynman theorem}
\label{subsec:case 3}
An independent method to obtain the spectrum of the Yukawa problem is to use the Hellmann-Feynman theorem \cite{Güttinger1932,Pauli1933,Hellmann1939,Feynman1939}. From the Hamiltonian Eq.~(\ref{eq:hamiltonian}), we get: 
\begin{equation}
\frac{\partial E_{n}}{\partial k}=\left\langle \psi _{n}\right\vert \frac{\partial H(k)}{\partial k}\left\vert \psi _{n}\right\rangle =Ze^{2}\left\langle e^{-kr}\right\rangle
\label{eq:HFT}
\end{equation}

Doing the integration over $k$, we obtain the energy as follows
\begin{equation}
E^{\mathrm{HFT}}_{n_r,l}(k)=E_{n_r,\ell}^{(0)}+\int_0^k\ev{e^{-k' r}}_{{U_{n_r,\ell}}(k')}dk'.
\label{eq:hft_energy}
\end{equation}
with $E^{(0)}_{n_r,\ell}=-\frac{ \mu (Ze^{2})^2}{2\hbar^2(n_r + \ell + 1)^2}$ because at $k=0$, the SCP gives exactly the Coulomb potential.

If we insert the approximate wavefunctions ${U_{n_r,\ell}}(k')$ from \ref{eq:pert-scp} into the expectation value $\left\langle e^{-kr}\right\rangle$, the integrand becomes a polynomial in $k'$ times exponentials. Integrating term by term and expanding in $k$ up the order $k^{3}$ gives exactly the same series \ref{eq:Coulomb_Energy_3} (The calculation is presented in Appendix \ref{app:Inspired}). The agreement between these two precedent method serves as a strong consistency check and justifies the use of the corrected BCH eigenfunctions.

Alternatively, one may keep the integral form as an exact representation which gives us a approximate analytical expression of the Yukawa eigenenergies. The values of these integrals are given in Tables~\ref{tab:comp_energy_n0},\ref{tab:comp_energy_n1} and labeled $E_{HFT2}$ and $E_{HFT3}$ ($HFT2$ when using the second order corrected BCH wavefunctions and $HFT3$ for the third order ones). The results yield values virtually identical to the direct expectation method ($E_{exp2}$ and $E_{exp3}$). The difference between $E_{HFT3}$ and $E_{exp3}$ are negligible for all considered states, with differences appearing only at the $\inf 10^{-5} Ryd$ level or smaller. This mutual consistency between the direct integration and the Hellmann-Feynman theorem methods, validates the use of the corrected BCH wavefunctions and confirms that both methods produce the same analytic representation of the spectrum.

\section{Thermal Properties}
\label{sec:properties}
We now study the thermodynamic behavior of the hydrogen-like system embedded in a plasma, using the energy spectrum obtained from the exact BCH solutions for the states $n_r =0$. Although our corrected BCH wavefunctions provide more accurate energies, the exact BCH expressions \ref{eq:energy1} is sufficiently simple to allow a closed-form evaluation of the partition function, which is our main goal here. The extension to other $n_r$ states follows the same pattern but is more involved. One can consider that each value of $n_r$ is a separate sub-system and the full partition function is the sum of all partitions functions of these sub-systems\cite{Vicente2021}. 

The canonical partition function for $n_r =0$ states:
\begin{equation}
Z=\sum\limits_{n_{r}}\sum\limits_{l}e^{-\beta E_{0,l}}\text{ \ \ };\beta =
\frac{1}{k_{B}T} 
\end{equation}
Where $k_{B}$ is the Boltzmann constant and $T$ is the absolute temperature.
Because the sum over $\ell$ cannot be performed in closed form, we use the Euler-Maclaurin summation formula
\begin{equation*}
\sum\limits_{\ell=0}^{\infty }f\left( \ell\right) =\int\limits_{0}^{\infty
}f\left( x\right) dx+\frac{1}{2}f\left( 0\right) -\sum\limits_{p=1}^{\infty }
\frac{B_{2p}}{\left( 2p\right) !}f^{\left( 2p-1\right) }\left( 0\right) 
\end{equation*}
where $B_{2p}$ are Bernoulli numbers, $f^{\left( 2p-1\right) }$ represents the derivative of order $\left( 2p-1\right) $. The leading integral term dominates at high temperature, while the lower-order corrections become important at low temperatures. The integral term is expressed as follows
\begin{widetext}
\begin{equation}
I=\sum\limits_{j=0}^{\infty }\frac{e^{-\frac{Ze^{2}}{\lambda k_{B}T}\left( 1+\sqrt{\frac{\hbar ^{2}}{12Ze^{2}\mu \lambda }}\right) }}{j!}\left( \frac{\mu\left( Ze^{2}\right) ^{2}}{2\hbar ^{2}k_{B}T}\right) ^{j} \left( \frac{1}{k_{B}T}\sqrt{\frac{\hbar ^{2}\left( Ze^{2}\right) }{3\mu \lambda ^{3}}}\right) ^{2j-1}\Gamma \left[ 1-2j,\left( \frac{1}{k_{B}T}\sqrt{\frac{\hbar
^{2}\left( Ze^{2}\right) }{3\mu \lambda ^{3}}}\right) \right]
\label{eq:A1thermal}
\end{equation}
\end{widetext}
The derivation involves expanding the exponential in the integrand and using the integral
\begin{equation*}
\int_{u}^{+\infty} x^{\nu-1} e^{-\mu x}\, dx 
= \mu^{-\nu}\, \Gamma(\nu,\mu u) 
\quad\text{with}u > 0,\ \operatorname{Re}(\mu) > 0
\end{equation*}
as well as the incomplete gamma function for the lower limit contributions. Keeping only terms up to order $\lambda ^{-3}$, we obtain after straightforward calculation, the simplified form
\begin{align}
Z\left( \lambda ,T\right)=&\bigg\{ \chi \left( 1-\frac{\nu }{T}\frac{1}{
\lambda }+\frac{\nu ^{2}}{2T^{2}}\frac{1}{\lambda ^{2}}+\left( \frac{27}{32}
\frac{T}{\delta }-1\right) \frac{\nu ^{3}}{6T^{3}}\frac{1}{\lambda ^{3}}
\right) \nonumber\\
&-\frac{\nu ^{3}}{720T^{3}}\frac{1}{\lambda ^{3}}\bigg\}e^{\frac{\delta }{T}},
\end{align}
where we have introduced the shorthand notations
\begin{equation*}
\chi =\frac{1}{2}-\frac{1}{5}\frac{\sigma }{T}+\frac{1}{20}\frac{\sigma ^{2}
}{T^{2}}-\frac{1}{90}\frac{\sigma ^{3}}{T^{3}},\text{ }\sigma =\frac{\mu
\left( Ze^{2}\right) ^{2}}{2k_{B}\hbar ^{2}}\text{and }\nu =\frac{Ze^{2}}{
k_{B}} 
\end{equation*}
 
We rewrite the partition function in approximate form of order two in $T$ (up to $\mathcal{O}(T^{-2})$)
\begin{equation}
Z\left( \lambda ,T\right) =\frac{1}{2}+\frac{\theta }{2T}+\left( \theta ^{2}+
\frac{\sigma ^{2}}{25}+\frac{9}{32}\frac{\nu ^{3}}{\sigma \lambda ^{3}}
\right) \frac{1}{4T^{2}}
\label{eq:partition}
\end{equation}
with $\theta =(3\sigma /5)-(\nu /\lambda)$

This expression can be tested; using the limit $\lambda ^{-}\rightarrow 0$, and we obtain the ordinary partition function.

From this partition function, we obtain the following thermodynamic quantities: the free energy $F$, the mean energy $U$, the specific heat $C$ and the entropy $S$
\begin{equation}
F=-k_{B}T\ln \left[ Z\right]
\label{eq:free energy}
\end{equation}
\begin{equation}
U=k_{B}T^{2}\frac{\partial \ln Z}{\partial T}=-2k_{B}T\left( 1-\frac{1+\frac{
\theta }{2T}}{2Z}\right) 
\label{eq:mean energy}
\end{equation}
\begin{equation}
C=\frac{\partial U}{\partial T}=-2k_{B}\left( 1-\frac{1+\frac{\theta }{T}}{2Z }\right) 
\label{eq:specific heat}
\end{equation}
\begin{equation}
S=-\left( \frac{\partial F}{\partial T}\right) _{V}=k_{B}\left(-2+\ln Z+
\frac{1+\frac{\theta }{2T}}{Z}\right) 
\label{eq:entropy}
\end{equation}
These expressions are valid for temperatures sufficiently high that the expansion \ref{eq:partition} is accurate. In the limit of vanishing screening ($\lambda \to \infty$), the potential reduces to the Coulomb case, then $\theta  \to 3\sigma /5$, and the partition function becomes
\begin{equation}
Z_{Coulomb}(T) =\frac{1}{2}+\frac{3\sigma }{10T}+\frac{2\sigma ^{2}}{25 T^2}
\label{eq:Coulomb_partition}
\end{equation}
which agrees with the standard result for a hydrogen atom in a thermal bath (up to the given order).

The expressions of the thermodynamic properties \ref{eq:free energy}-\ref{eq:entropy} capture the leading thermal behavior and clearly exhibit the influence of plasma screening through the parameter $\lambda$. Graphical representations of these formulas is presented in the next section \ref{sec:results} (Figures \ref{fig:thermal1}-\ref{fig:thermal4}).

\section{ Results and Discussion}
\label{sec:results}
We now present the numerical results obtained from all three methods presented here: the exact BCH spectrum (column $E_{BCH}$ in Tables~\ref{tab:comp_energy_n0},\ref{tab:comp_energy_n1}) and the expectation values using the corrected BCH eigenfunctions from both the full Yukawa Hamiltonian (columns $E_{exp2}$ and $E_{exp3}$) and the Hellmann-Feynman theorem (columns $E_{HFT2}$ and $E_{HFT3}$). For comparison, we adopt the variational results of Paul and Ho work\cite{Paul2009} as the reference values. Their work, which employs a Ritz variational method with Coulomb wavefunctions, is itself an update of the accurate numerical benchmarks established by Rogers et al.\cite{Rogers1970}. These reference values are routinely used to validate all methods used for the Yukawa potential\cite{Gonul2006,Soylu2012,Hamzavi2012,Napsuciale2021}.

\begin{figure}[h]
\centering
\includegraphics[width=0.45\textwidth]{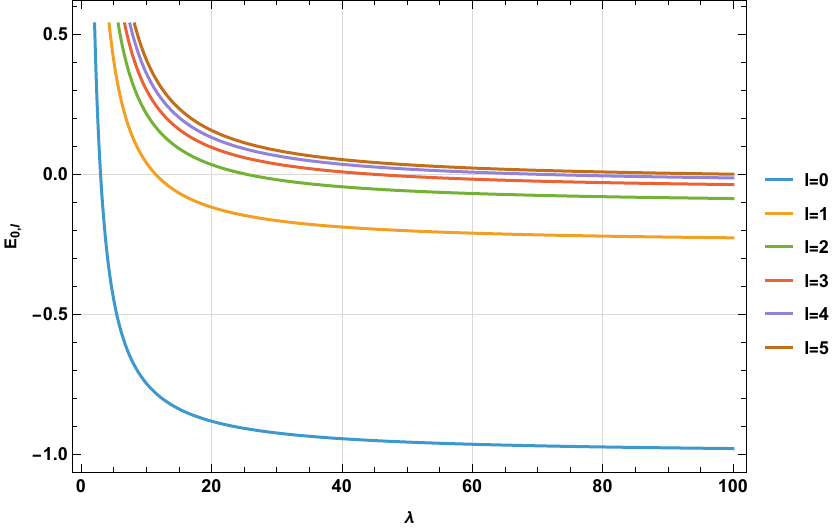}
\caption{The energy eigenvalues (in Rydberg units) of the hydrogen-like atomic systems immersed in plasma medium as a function of different screening parameters $\left( \lambda \right) $ in the case $\left( n_{r}=0\right) $.}
\label{fig:fig1}
\end{figure}

\begin{figure}[h]
\centering
\includegraphics[width=0.45\textwidth]{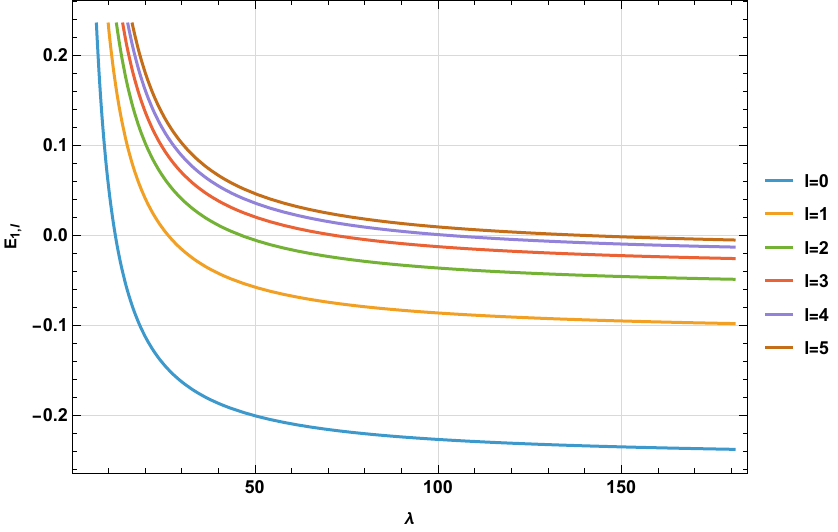}
\caption{The energy eigenvalues (in Rydberg units) of the hydrogen-like atomic systems immersed in plasma medium as a function of different screening parameters $\left( \lambda \right) $ in the case $\left( n_{r}=1\right) $.}
\label{fig:fig2}
\end{figure}

The exact BCH energies ($E_{BCH}$) agree with the reference values for weak screening (large $\lambda_D$) but deviate for large screening for the $1s$ state and for medium screening when considering the other states. For example, for the $1s$ state at $\lambda_D=20$, the relative error is about $2.54\%$, and it reaches $8.46\%$ at $\lambda_D=10$. This confirms that the validity of exact polynomial truncation is limited at weak screening and does not provide a smooth analytic continuation to the Coulomb limit. The behavior of exact BCH energies is shown in Figures \ref{fig:fig1} and \ref{fig:fig2}.

In stark contrast, the analytic corrected BCH wavefunctions, already at second order ($E_{exp2}$), give energies much closer to the reference values. Including the third-order corrections ($E_{exp3}$) reduces the errors to below $0.01\%$ for most values with $\lambda_D \gtrsim 20$. For instance, for the $1s$ state at $\lambda_D = 20$, $\epsilon{exp3-P}=0.0003\%$, and at $\lambda_D = 20$, it is only $0.0019\%$. Similar improvements are seen for the other states; this demonstrates the analytic convergence of our BCH inspired series.

The Hellmann-Feynman method yields results virtually identical to the direct expectation values; differences between $E_{HFT3}$ and $E_{exp3}$ are typically at the $10^{-5}Ry$ level or smaller, providing a strong internal consistency check. It is important to emphasize that the H-F theorem is an exact relation that holds for the true eigenfunctions of the Hamiltonian. When applied with approximate wavefunctions, the degree to which the integrated energy reproduces the result obtained from direct expectation method provides a sensitive measure of the quality of those wavefunctions. In our case, the close agreement between $E_{HFT}$ and $E_{ex}$ in not guaranteed a priori; it demonstrates that our BCH-inspired wavefunctions are sufficiently accurate to satisfy the H-F theorem to the order considered. Thus the consistency between these two independent methods and their accuracy regarding the benchmarks results, serve as a powerful validation of the BCH-inspired wavefunctions themselves, beyond the mere reproduction of the energy eigenvalues.

Importantly, our method also provides closed-form analytic expressions beyond the power series. For example, the energies for states where $n_r =0$ are given in both Appendices \label{app:Full_Exp_Yukawa} and \label{app:HFT} as an elementary function of $k$. For higher states, the results reduce to finite sum of hypergeometric functions; such closed forms are not available in purely numerical or variational approaches.

\begin{figure*}[t]
    \centering
    \includegraphics[width=0.3\textwidth]{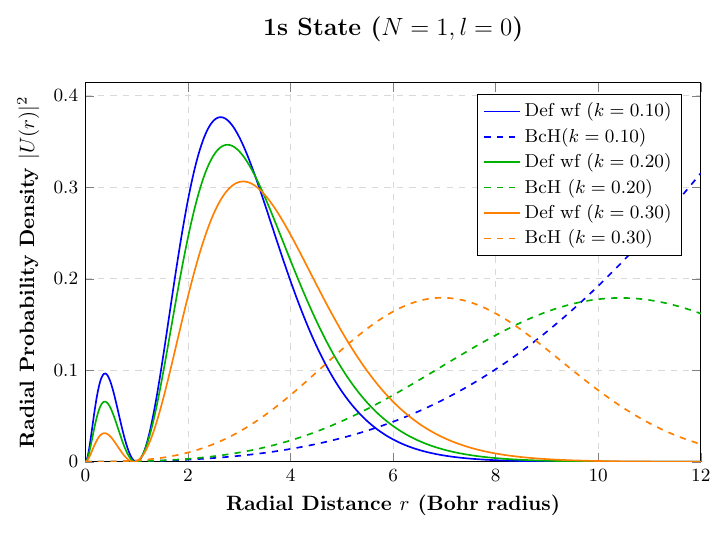}
    \includegraphics[width=0.3\textwidth]{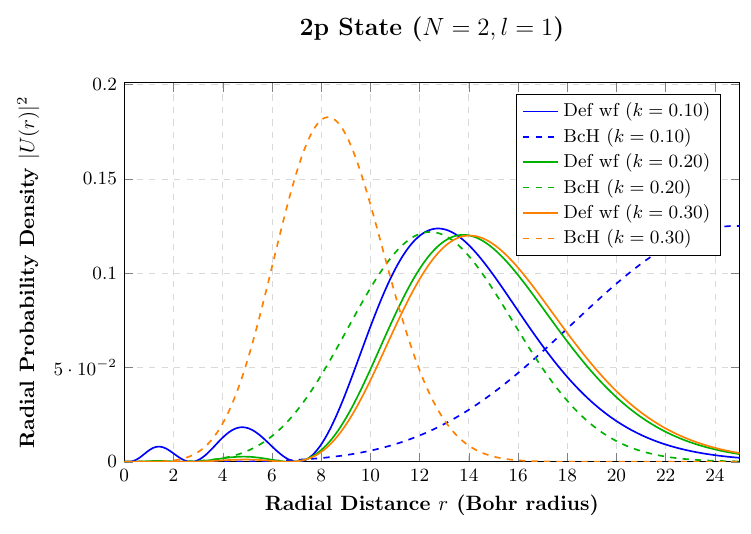}
    \includegraphics[width=0.3\textwidth]{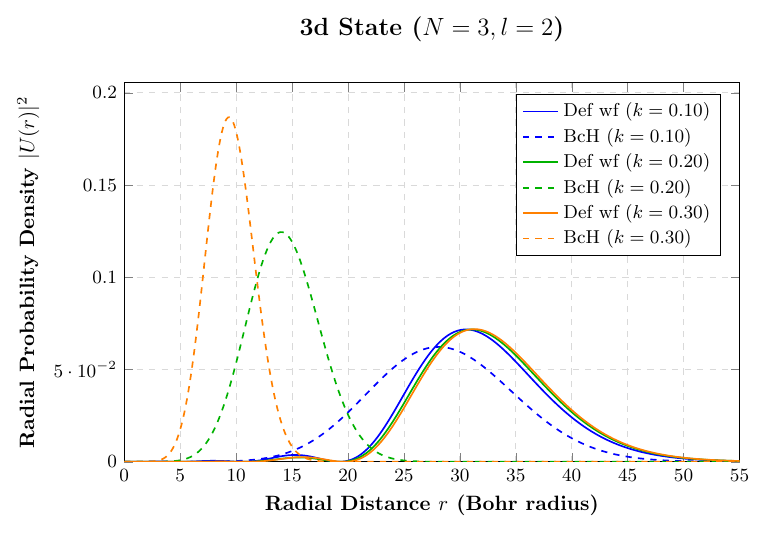}
    \caption{Normalized radial probability distributions $P_{n_{r},\ell}\left( r\right) =r^{2}\left\vert R_{n_{r},\ell}\right\vert ^{2}$ for the$(n_{r}=0,\ell=0),(n_{r}=0,l=1)$ and $(n_{r}=0,\ell=2)$ states as functions of radial length for various screening parameters $\left(k=1/\lambda_{D} \right)$.}
    \label{fig:wavefunctions n=0}
\end{figure*}

\begin{figure*}[t]
    \centering
    \includegraphics[width=0.3\textwidth]{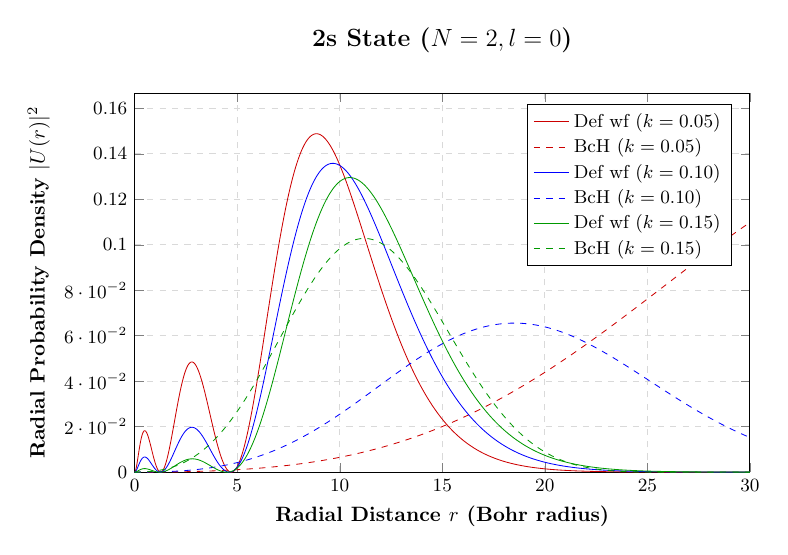}
    \includegraphics[width=0.3\textwidth]{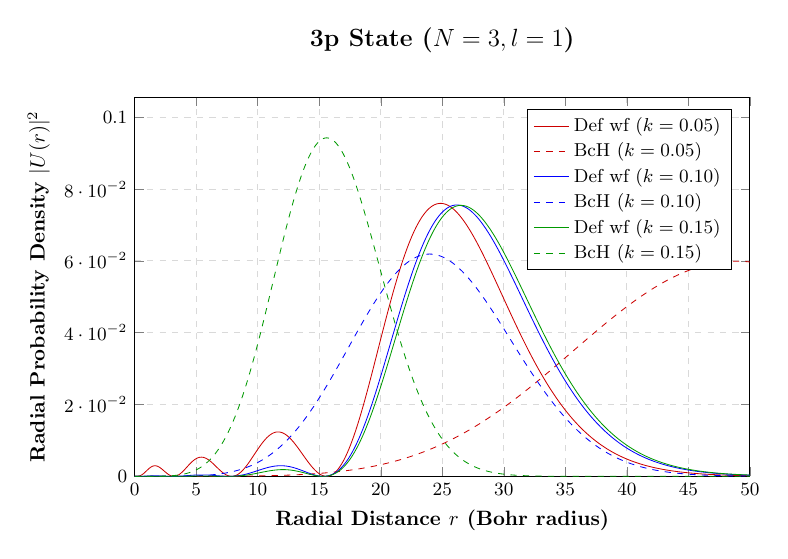}
    \includegraphics[width=0.3\textwidth]{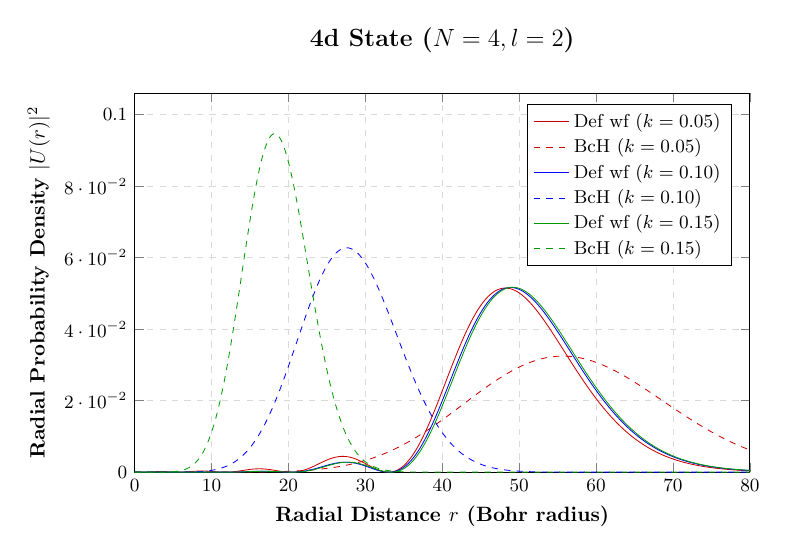}
    \caption{Normalized radial probability distributions $P_{n_{r},\ell}\left( r\right) =r^{2}\left\vert R_{n_{r},\ell}\right\vert ^{2}$ for the$(n_{r}=1,\ell=0),(n_{r}=1,\ell=1)$ and $(n_{r}=1,\ell=2)$ states as functions of radial length for various screening parameters  $\left( k=1/\lambda_{D} \right)$.}
    \label{fig:wavefunctions n=1}
\end{figure*}

\begin{figure}[t]
\centering
\includegraphics[width=0.45\textwidth]{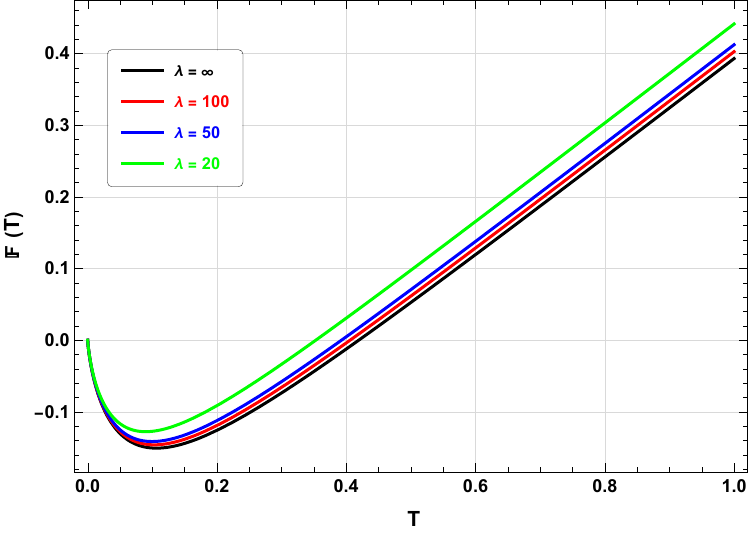}
\caption{The free energy $F$ as a function of temperature$T$ for different values of $\left( \lambda \right) $.\\
{( $T_{min}=0.089671$,$ T_{0}=0.352838 $ for $ \lambda=20$),($T_{min}=0.0997185$,$ T_{0}=0.392591 $ for $ \lambda=50$),($T_{min}=0.103088$,$ T_{0}=0.405912 $ for $ \lambda=100$),($T_{min}=0.106464$,$ T_{0}=0.419258 $ for $ \lambda=\infty$)}}
\label{fig:thermal1}
\end{figure}

\begin{figure}[t]
\centering
\includegraphics[width=0.45\textwidth]{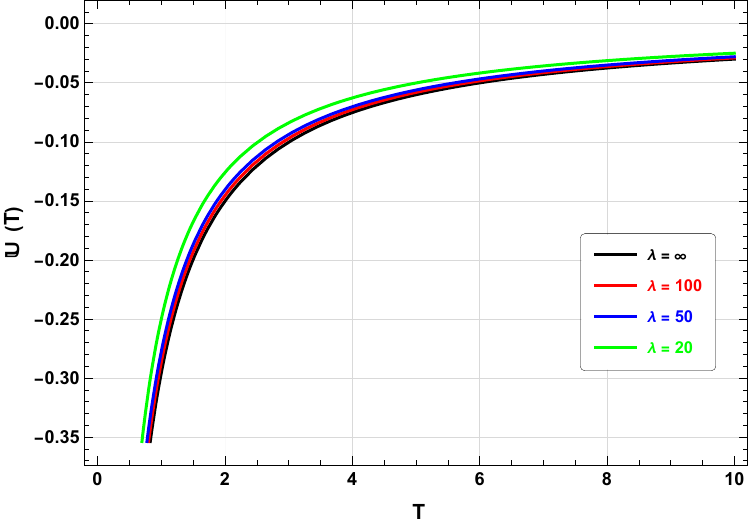}
\caption{The mean energy $U$ as a function of temperaturev$T$ for different values of $\left( \lambda \right) $.}
\label{fig:thermal2}
\end{figure}

\begin{figure}[t]
\centering
\includegraphics[width=0.45\textwidth]{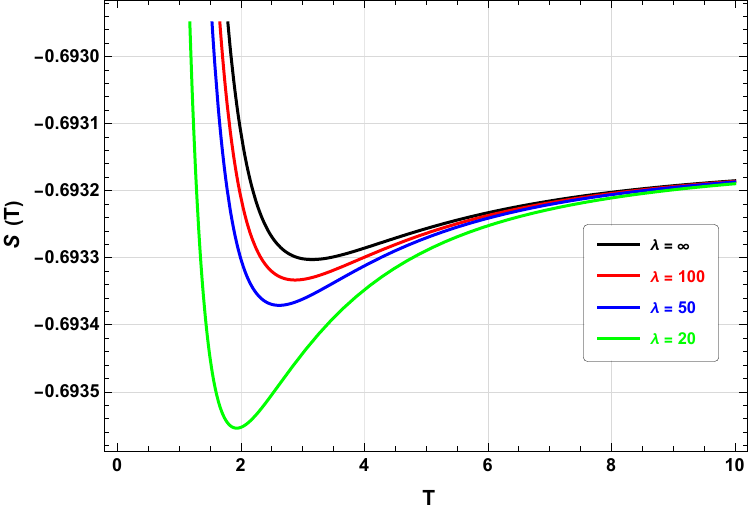}
\caption{The entropy $S$ as a function of temperaturev$T$ for different values of $\left( \lambda \right) $.\\
{( $T_{min}=1.93661$ for $ \lambda=20$),($T_{min}=2.62312$ for $ \lambda=50$),($T_{min}=2.88236$ for $ \lambda=100$),($T_{min}=3.15831$ for $ \lambda=\infty$)}}
\label{fig:thermal3}
\end{figure}

\begin{figure}[t]
\centering
\includegraphics[width=0.45\textwidth]{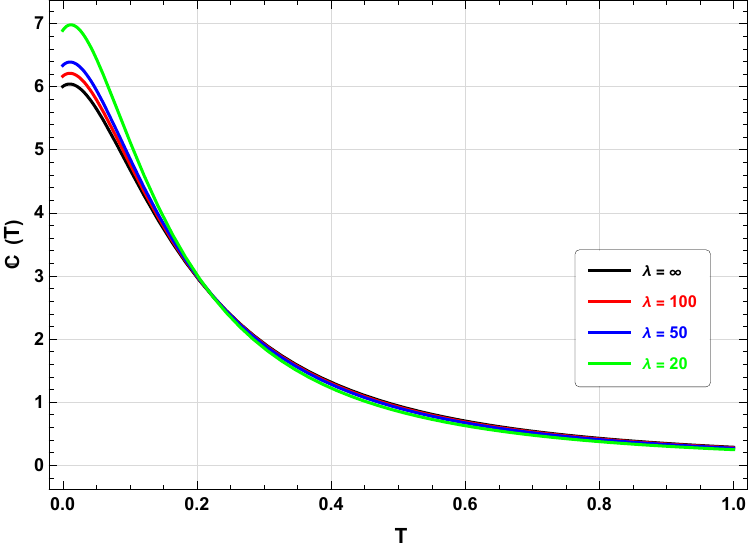}
\caption{Specific heat $C$ as a function of temperaturev$T$ for different values of $\left( \lambda \right) $.}
\label{fig:thermal4}
\end{figure}

The radial probability distribution Figures~\ref{fig:wavefunctions n=0}-\ref{fig:wavefunctions n=1} were computed directly from our analytic wavefunctions. As screening increases, the distribution broaden, reflecting the reduced nuclear attraction. The effect is more pronounced for higher $\ell$ states.

The thermodynamic properties derived from the partition function (Figures \ref{fig:thermal1}-\ref{fig:thermal4}) were also obtained analytically using the Euler-Maclaurin summation formula. The free energy $F$ decreases rapidly to a minimum then increases. The mean energy $U$ increases monotonically with temperature. The entropy $S$ and the specific heat $C$ exhibit characteristic behaviors that depends on the screening length. All thermodynamic functions converges to the pure Coulomb limits when $\lambda \to \infty$.


\section{Conclusions}
\label{sec:conclusion}
In this article, we have developed a fully analytical methods to study plasma screening effects on hydrogen-like atomic systems described by the screened Coulomb (Yukawa) potential. Starting from the detailed analytic analysis of the Schrödinger equation for the cubic-truncated Yukawa potential based on the biconfluent Heun differential equation. The energy eigenvalues and corresponding wave functions of the system are obtained by applying the full boundary conditions on the solutions.

Because these eigenfunctions do not have the Coulomb ones as asymptotic solutions for vanishing screening, we abandoned the rigid polynomial truncation relations, and use their form to write inspired bi-confluent Heun eigenfunctions that are analytic in $k$ and reduce exactly to Coulomb wavefunctions when $k=0$. The construction is systematic and we provided explicit formulas up to $\mathcal{O}(k^3)$ and indicated how to extend to higher order.

Using these inspired BCH wavefunctions, we computed the energy spectrum via two independent analytic methods: direct expectation value of the full Yukawa Hamiltonian giving closed-form hypergeometric expressions and the Hellmann-Feynman theorem yielding an integral representation. Both methods reproduce the Rayleigh-Schrödinger perturbative expansion and agree perfectly with each other. Numerical results for $n_r =0$ and $n_r =1$ states show excellent agreement with reference data for a wide range of screening lengths, with errors below $0.01\%$ for most states when third-order corrections are included. The mutual agreement between both direct expectation value and Hellmann-Feynman theorem methods - an exact quantum mechanical relation - confirms that our corrected BCH wavefunctions are not only energy-accurate but also faithfully represent the true eigenfunctions in the sense required by the Hellmann-Feynman theorem. 

We also studied the radial probability distributions and the thermodynamic properties and all quantities are expressed analytically and exhibit the expected physical behavior.

The analytic nature of our approach offers significant advantages over purely numerical and variational methods: explicit formulas, systematic improvement with higher order, direct calculations and smooth Coulomb limit. It is worth emphasizing that the exact BCH solutions already represent the closest possible analytic solutions to the truncated Yukawa potential, because no analytic solution exists when an $r^3$ term is included in the potential. Moreover both exact BCH form and their corrected inspired wavefunctions are ideally suited for variational calculations: the exact form contains an exponential factor $\text{exp}(-\beta r -\gamma r^2)$ with two possible variational parameters, while the inspired form is a finite sum of Laguerre polynomials times a Coulomb exponential, offering a flexible, systematically improvable trial function. Thus beyond providing perturbative energies, the wavefunctions presented can serve as a starting point for variational studies with one or more parameters, further extending their utility.

The present work establishes a powerful analytic framework for studying confined systems with screened interactions. It can be easily extended to ,other exponential-type potentials, and to relativistic equations.

\begin{acknowledgments}
This work was supported by PRFU B00L02UN050120230005 of the DGRSDT (Algeria).
\end{acknowledgments}

\begin{table*}[t]
\centering
\scriptsize

\caption{
Comparison of the energy eigenvalues (in Rydberg) for the
$n_r=0,\ell$ states obtained using different methods, together with
their relative errors with respect to the reference values
$E_{\rm Paul}$ for various Debye lengths $\lambda_D$.
The relative error is defined by
$\epsilon = 100\,|E-E_{\rm Paul}|/|E_{\rm Paul}|$ (\%).
}

\label{tab:comp_energy_n0}

\renewcommand{\arraystretch}{1.1}
\setlength{\tabcolsep}{1.5pt}

\begin{tabular*}{\linewidth}{@{\extracolsep{\fill}}
c c
r r r r r r r
r r r
@{}}

\toprule

State &
$\lambda_D$
&
$E_{\rm Paul}$
&
$E_{\rm exp2}$
&
$E_{\rm exp3}$
&
$E_{\rm HFT2}$
&
$E_{\rm HFT3}$
&
$E_{\rm BCH}$
&
$\epsilon_{\rm exp3-P}$
&
$\epsilon_{\rm HFT3-P}$
&
$\epsilon_{\rm BCH-P}$
\\
\midrule

\multirow{9}{*}{1s}

 & $200$ & $-0.9900370$ & $-0.9900404$ & $-0.9900404$ & $-0.9900374$ & $-0.9900374$ & $-0.9893876$ & $0.0003$ & $0.0000$ & $0.0656$ \\
 & $150$ & $-0.9867330$ & $-0.9867360$ & $-0.9867360$ & $-0.9867330$ & $-0.9867330$ & $-0.9857239$ & $0.0003$ & $0.0000$ & $0.1023$ \\
 & $100$ & $-0.9801490$ & $-0.9801520$ & $-0.9801520$ & $-0.9801490$ & $-0.9801490$ & $-0.9782679$ & $0.0003$ & $0.0000$ & $0.1919$ \\
 & $70$ & $-0.9717320$ & $-0.9717348$ & $-0.9717348$ & $-0.9717318$ & $-0.9717318$ & $-0.9684711$ & $0.0003$ & $0.0000$ & $0.3356$ \\
 & $50$ & $-0.9605920$ & $-0.9605951$ & $-0.9605951$ & $-0.9605921$ & $-0.9605921$ & $-0.9551010$ & $0.0003$ & $0.0000$ & $0.5716$ \\
 & $40$ & $-0.9509220$ & $-0.9509251$ & $-0.9509252$ & $-0.9509221$ & $-0.9509222$ & $-0.9431535$ & $0.0003$ & $0.0000$ & $0.8169$ \\
 & $30$ & $-0.9349640$ & $-0.9349667$ & $-0.9349671$ & $-0.9349637$ & $-0.9349639$ & $-0.9227924$ & $0.0003$ & $0.0000$ & $1.3018$ \\
 & $20$ & $-0.9036320$ & $-0.9036318$ & $-0.9036345$ & $-0.9036288$ & $-0.9036302$ & $-0.8806351$ & $0.0003$ & $0.0002$ & $2.5449$ \\
 & $10$ & $-0.8141030$ & $-0.8140620$ & $-0.8141184$ & $-0.8140590$ & $-0.8140988$ & $-0.7452277$ & $0.0019$ & $0.0005$ & $8.4603$ \\

\midrule

\multirow{9}{*}{2p}

 & $200$ & $-0.2401240$ & $-0.2401238$ & $-0.2401238$ & $-0.2401238$ & $-0.2401238$ & $-0.2389794$ & $0.0001$ & $0.0001$ & $0.4767$ \\
 & $150$ & $-0.2368860$ & $-0.2368860$ & $-0.2368860$ & $-0.2368860$ & $-0.2368860$ & $-0.2350953$ & $0.0000$ & $0.0000$ & $0.7559$ \\
 & $100$ & $-0.2304900$ & $-0.2304902$ & $-0.2304902$ & $-0.2304902$ & $-0.2304902$ & $-0.2271132$ & $0.0001$ & $0.0001$ & $1.4650$ \\
 & $70$ & $-0.2224210$ & $-0.2224206$ & $-0.2224209$ & $-0.2224205$ & $-0.2224207$ & $-0.2164995$ & $0.0001$ & $0.0001$ & $2.6623$ \\
 & $50$ & $-0.2119260$ & $-0.2119227$ & $-0.2119243$ & $-0.2119227$ & $-0.2119234$ & $-0.2018350$ & $0.0008$ & $0.0012$ & $4.7616$ \\
 & $40$ & $-0.2029830$ & $-0.2029753$ & $-0.2029799$ & $-0.2029753$ & $-0.2029774$ & $-0.1885891$ & $0.0015$ & $0.0028$ & $7.0912$ \\
 & $30$ & $-0.1885630$ & $-0.1885390$ & $-0.1885564$ & $-0.1885390$ & $-0.1885473$ & $-0.1657651$ & $0.0035$ & $0.0083$ & $12.0903$ \\
 & $20$ & $-0.1614610$ & $-0.1613513$ & $-0.1614571$ & $-0.1613512$ & $-0.1614084$ & $-0.1177251$ & $0.0024$ & $0.0326$ & $27.0876$ \\
 & $10$ & $-0.0928450$ & $-0.0915067$ & $-0.0930425$ & $-0.0915067$ & $-0.0928683$ & $0.0412871$ & $0.2128$ & $0.0251$ & $144.4688$ \\

\midrule

\multirow{9}{*}{3d}

& $200$ & $-0.1013690$ & $-0.1013684$ & $-0.1013685$ & $-0.1013684$ & $-0.1013685$ & $-0.0996822$ & $0.0005$ & $0.0005$ & $1.6640$ \\
 & $150$ & $-0.0982330$ & $-0.0982323$ & $-0.0982324$ & $-0.0982323$ & $-0.0982323$ & $-0.0955779$ & $0.0007$ & $0.0007$ & $2.7029$ \\
 & $100$ & $-0.0921230$ & $-0.0921205$ & $-0.0921211$ & $-0.0921205$ & $-0.0921207$ & $-0.0870697$ & $0.0021$ & $0.0025$ & $5.4854$ \\
 & $70$ & $-0.0845740$ & $-0.0845658$ & $-0.0845691$ & $-0.0845658$ & $-0.0845672$ & $-0.0756390$ & $0.0058$ & $0.0080$ & $10.5647$ \\
 & $50$ & $-0.0750260$ & $-0.0749965$ & $-0.0750128$ & $-0.0749965$ & $-0.0750038$ & $-0.0596802$ & $0.0176$ & $0.0296$ & $20.4540$ \\
 & $40$ & $-0.0671380$ & $-0.0670688$ & $-0.0671140$ & $-0.0670688$ & $-0.0670899$ & $-0.0451359$ & $0.0358$ & $0.0716$ & $32.7715$ \\
 & $30$ & $-0.0549120$ & $-0.0547145$ & $-0.0548751$ & $-0.0547145$ & $-0.0547954$ & $-0.0198490$ & $0.0672$ & $0.2123$ & $63.8532$ \\
 & $20$ & $-0.0337270$ & $-0.0328803$ & $-0.0337197$ & $-0.0328803$ & $-0.0333803$ & $0.0340737$ & $0.0217$ & $1.0280$ & $201.0279$ \\
 & $10$ & --- & $0.0150383$ & $0.0067784$ & $0.0150383$ & $0.0082556$ & $0.2166908$ & --- & --- & --- \\

\bottomrule

\end{tabular*}
\end{table*}

\begin{table*}[t]
\centering
\scriptsize

\caption{
Comparison of the energy eigenvalues (in Rydberg) for the
$n_r=1,\ell$ states obtained using different methods, together with
their relative errors with respect to the reference values
$E_{\rm Paul}$ for various Debye lengths $\lambda_D$.
The relative error is defined by
$\epsilon = 100\,|E-E_{\rm Paul}|/|E_{\rm Paul}|$ (\%).
}

\label{tab:comp_energy_n1}

\renewcommand{\arraystretch}{1.1}
\setlength{\tabcolsep}{1.5pt}

\begin{tabular*}{\linewidth}{@{\extracolsep{\fill}}
c c
r r r r r r r
r r r
@{}}

\toprule

State &
$\lambda_D$
&
$E_{\rm Paul}$
&
$E_{\rm exp2}$
&
$E_{\rm exp3}$
&
$E_{\rm HFT2}$
&
$E_{\rm HFT3}$
&
$E_{\rm BCH}$
&
$\epsilon_{\rm exp3-P}$
&
$\epsilon_{\rm HFT3-P}$
&
$\epsilon_{\rm BCH-P}$
\\
\midrule

\multirow{9}{*}{2s}
 & $200$ & $-0.2401480$ & $-0.2401487$ & $-0.2401487$ & $-0.2401483$ & $-0.2401483$ & $-0.2387754$ & $0.0003$ & $0.0001$ & $0.5716$ \\
 & $150$ & $-0.2369290$ & $-0.2369297$ & $-0.2369297$ & $-0.2369292$ & $-0.2369293$ & $-0.2347813$ & $0.0003$ & $0.0001$ & $0.9065$ \\
 & $100$ & $-0.2305870$ & $-0.2305867$ & $-0.2305868$ & $-0.2305863$ & $-0.2305864$ & $-0.2265369$ & $0.0001$ & $0.0003$ & $1.7564$ \\
 & $70$ & $-0.2226150$ & $-0.2226138$ & $-0.2226144$ & $-0.2226134$ & $-0.2226140$ & $-0.2155166$ & $0.0003$ & $0.0005$ & $3.1886$ \\
 & $50$ & $-0.2122970$ & $-0.2122928$ & $-0.2122957$ & $-0.2122923$ & $-0.2122956$ & $-0.2002100$ & $0.0006$ & $0.0007$ & $5.6934$ \\
 & $40$ & $-0.2035530$ & $-0.2035422$ & $-0.2035505$ & $-0.2035416$ & $-0.2035514$ & $-0.1863224$ & $0.0012$ & $0.0008$ & $8.4649$ \\
 & $30$ & $-0.1895470$ & $-0.1895147$ & $-0.1895452$ & $-0.1895136$ & $-0.1895534$ & $-0.1622880$ & $0.0009$ & $0.0034$ & $14.3811$ \\
 & $20$ & $-0.1635580$ & $-0.1634120$ & $-0.1635807$ & $-0.1634085$ & $-0.1636823$ & $-0.1113920$ & $0.0139$ & $0.0760$ & $31.8945$ \\
 & $10$ & $-0.1000580$ & $-0.0983944$ & $-0.0999701$ & $-0.0983482$ & $-0.1040381$ & $0.0586115$ & $0.0878$ & $3.9778$ & $158.5775$ \\
\midrule

\multirow{9}{*}{3p}
 & $200$ & $-0.1014160$ & $-0.1014163$ & $-0.1014163$ & $-0.1014163$ & $-0.1014163$ & $-0.0994109$ & $0.0003$ & $0.0003$ & $1.9771$ \\
 & $150$ & $-0.0983170$ & $-0.0983160$ & $-0.0983162$ & $-0.0983160$ & $-0.0983162$ & $-0.0951608$ & $0.0008$ & $0.0008$ & $3.2102$ \\
 & $100$ & $-0.0923060$ & $-0.0923035$ & $-0.0923048$ & $-0.0923034$ & $-0.0923049$ & $-0.0863064$ & $0.0013$ & $0.0012$ & $6.4996$ \\
 & $70$ & $-0.0849360$ & $-0.0849254$ & $-0.0849330$ & $-0.0849254$ & $-0.0849338$ & $-0.0743436$ & $0.0036$ & $0.0026$ & $12.4710$ \\
 & $50$ & $-0.0757060$ & $-0.0756674$ & $-0.0757034$ & $-0.0756672$ & $-0.0757098$ & $-0.0575542$ & $0.0034$ & $0.0051$ & $23.9767$ \\
 & $40$ & $-0.0681620$ & $-0.0680732$ & $-0.0681710$ & $-0.0680729$ & $-0.0681954$ & $-0.0421918$ & $0.0131$ & $0.0490$ & $38.1007$ \\
 & $30$ & $-0.0566280$ & $-0.0563791$ & $-0.0567104$ & $-0.0563783$ & $-0.0568372$ & $-0.0153922$ & $0.1455$ & $0.3694$ & $72.8188$ \\
 & $20$ & $-0.0371650$ & $-0.0361499$ & $-0.0376780$ & $-0.0361459$ & $-0.0386619$ & $0.0419555$ & $1.3803$ & $4.0278$ & $212.8898$ \\
 & $10$ & $-0.0037210$ & $0.0060191$ & $-0.0044910$ & $0.0060755$ & $-0.0117919$ & $0.2362857$ & $20.6944$ & $216.9012$ & $6450.0583$ \\
\midrule

\multirow{9}{*}{4d}
 & $200$ & $-0.0530060$ & $-0.0530047$ & $-0.0530050$ & $-0.0530047$ & $-0.0530050$ & $-0.0503593$ & $0.0019$ & $0.0019$ & $4.9932$ \\
 & $150$ & $-0.0500550$ & $-0.0500523$ & $-0.0500536$ & $-0.0500523$ & $-0.0500536$ & $-0.0458729$ & $0.0028$ & $0.0027$ & $8.3550$ \\
 & $100$ & $-0.0444560$ & $-0.0444425$ & $-0.0444512$ & $-0.0444425$ & $-0.0444519$ & $-0.0364580$ & $0.0107$ & $0.0091$ & $17.9908$ \\
 & $70$ & $-0.0378160$ & $-0.0377674$ & $-0.0378131$ & $-0.0377674$ & $-0.0378191$ & $-0.0236348$ & $0.0077$ & $0.0081$ & $37.5005$ \\
 & $50$ & $-0.0298800$ & $-0.0297161$ & $-0.0299186$ & $-0.0297159$ & $-0.0299601$ & $-0.0055043$ & $0.1293$ & $0.2682$ & $81.5786$ \\
 & $40$ & $-0.0237400$ & $-0.0233809$ & $-0.0238897$ & $-0.0233806$ & $-0.0240263$ & $0.0111796$ & $0.6304$ & $1.2058$ & $147.0919$ \\
 & $30$ & $-0.0150650$ & $-0.0140994$ & $-0.0155733$ & $-0.0140985$ & $-0.0160935$ & $0.0404327$ & $3.3742$ & $6.8272$ & $368.3884$ \\
 & $20$ & $-0.0031430$ & $0.0005987$ & $-0.0041906$ & $0.0006031$ & $-0.0060824$ & $0.1034376$ & $33.3313$ & $93.5216$ & $3391.0460$ \\
 & $10$ & --- & $0.0255651$ & $0.0097522$ & $0.0256228$ & $0.0056858$ & $0.3192492$ & --- & --- & --- \\

\bottomrule

\end{tabular*}
\end{table*}

\appendix
\section{Normalization Constants}
\label{app:Norms-BCH}
The radial functions are normalized through
\begin{equation*}
	\int_0^\infty |U_{n\ell}(r)|^2\,dr=1, \qquad U_{n\ell}(r)=rR_{n\ell}(r).
\end{equation*}
We use the following integral formula from Table of Integrals, Series, and Products, p. 365 \cite{Gradshteyn2007}
\begin{equation*}
\int_{0}^{\infty} x^{\nu-1} e^{-a x^{2}-b x}\,dx=
(2a)^{\frac{-\nu}{2}}\Gamma(\nu)
\exp\!\left(\frac{b^{2}}{8a}\right)
D_{-\nu}\!\left(\frac{b}{\sqrt{2a}}\right),
\end{equation*}
\[
\qquad
\Re(a)>0,\ \Re(\nu)>0,
\]
where $D_{-\upsilon}$ represents the parabolic cylinder function.

After integrating the radial functions (from Eqs.~(\ref{eq:wave function1}) and (\ref{eq:wave function2})) and simplifying, we obtain the following normalization constants:

\subsection{Case $\left( n_{r}=0\right)$ of the BCH Solutions}
\label{subsec:case 1}
\begin{equation}
	A_{0,\ell}=\sqrt{\frac{2(3\lambda ^{2})^{-\left( \ell+\frac{3}{2}\right)
			}\sigma ^{\ell+\frac{3}{2}}}{\left( \sqrt{3\sigma }\Gamma (\ell+2)F_{\frac{1}{2}
			}^{\backprime }+\Gamma \left( \ell+\frac{3}{2}\right) F_{1}\right) }}
	\label{eq:A1}
\end{equation}
\subsection{Case $\left( n_{r}=1\right)$ of the BCH Solutions}
\label{subsec:case 2}
\begin{equation}
	{\footnotesize A_{1,\ell}=\sqrt{\frac{\sqrt{3}\left( 2\hbar \right) ^{2}\left( \frac{2\sigma }{3\lambda ^{2}}\right) ^{l+\frac{3}{2}}\Gamma (2\ell+3)^{2}}
			{\mathcal{D}}}}
	\label{eq:A2}
\end{equation}
Where
\[
\mathcal{D}=\begin{array}{c}
	\Gamma (2\ell+2)\left( \sigma +\frac{3}{2}(\ell+1)\right)  \\ 
	\begin{array}{c}
		\left[\sigma \Gamma (2\ell+2)\left( \sigma +\frac{3}{2}(\ell+1)\right) \left( \Gamma \left(\ell+\frac{5}{2}\right) F_{2}+\sqrt{
			6\sigma }\Gamma (\ell+3)\,F_{\frac{3}{2}}^{\backprime }\right) \right.  \\ 
		\left. -\hbar \Gamma (2\ell+3)\left( 2\sqrt{6\sigma }\Gamma (\ell+2)\,F_{1}^{\backprime
		}+6\sigma \Gamma \left(\ell+\frac{5}{2}\right) \,F_{2}\right) \right]
	\end{array}
	\\ 
	+6\hbar ^{2}\Gamma (2\ell+3)^{2}
	\left( \Gamma \left(\ell+\frac{3}{2}\right)\,F_{1}+\sqrt{6\sigma }\Gamma (\ell+2)\,F_{\frac{1}{2}}^{\backprime }\right) 
\end{array}
\]
\[
\sigma {\small =}\sqrt{\frac{3\lambda \mu Ze^{2}}{\hbar ^{2}}}
\]
\[
F_{q}=_{1}F_{1}\left( l+q+\frac{1}{2};\frac{1}{2};\frac{3\sigma }{4}\right) 
\]
And
\[
F_{q}^{\backprime }=_{1}F_{1}\left( \ell+q+\frac{3}{2};\frac{3}{2};\frac{3\sigma }{4}\right) 
\]

\subsection{Corrected BCH Solutions}
\label{subsec:perturbative}
We use the corrected radial function is Eq.~(\ref{eq:pert-scp}) in the normalization condition, and expand to write:
\begin{align}
	& \int |U|^2_{n_r,\ell} dr=\mathcal N_{n_r,\ell}^2 \int r^{2\ell+2} e^{-2r/a_0(n_r+\ell+1)}\\
	&\times \Big[
	L^2_{n_r} +2k^2 L_{n_r} P^{(2)}_{n_r,\ell} +2k^3 L_{n_r} P^{(3)}_{n_r,\ell} +k^4 (P^{(2)})^2 \Big] dr,\nonumber
	\label{eq:Cor_BCH_Int}
\end{align}
We use $x = 2r/a_0 n$ and Laguerre orthogonality relations:
\begin{equation*}
	\int_0^\infty
	x^{\alpha}
	e^{-x}
	L_n^{\alpha}(x)
	L_m^{\alpha}(x)
	dx=\frac{\Gamma(n+\alpha+1)}{n!}\delta_{nm},
\end{equation*}
with
\begin{equation*}
	\int L_{n_r} L_{n_r-1} = 0,
	\qquad
	\int L_{n_r} L_{n_r-2} = 0,
\end{equation*}
To get the following expressions:
\begin{align}
	& \mathcal N_{n,\ell}=\sqrt{\frac{2}{a_0 n^2}\frac{(n-\ell-1)!}{n!}}\nonumber \\
	&\times
	\Bigg[1-\frac{k^2 a_0^2}{8}\left(3n^2-\ell(\ell+1)\right)\nonumber\\
	&-\frac{k^3 a_0^3n^2}{24}\left(5n^2+1-3\ell(\ell+1)\right)\Bigg]. 
	\label{eq:norm_cor_BCH3}
\end{align}
where we have used the Coulomb convention for the quantum numbers $n=n_r+\ell+1$. Detailed computations up to $k^2$ are given in the following Appendix.

\section{Detailed Construction of the Corrected BCH Wavefunctions}
\label{app:Inspired}
We start frpm the BCH ansatz
\begin{equation}
U_{n_r,l}(r)=r^{\ell+1}e^{-\alpha r-\beta r^2}g(r), \text{ } g(r)=\sum_{j=0}^{n_r}a_jr^j,
\tag{B.1}
\label{eq:Ansatz_Cor_BCH}
\end{equation}
with the physical requirement that at the Coulomb limit $k\to 0$, we have
\begin{equation}
\alpha\to\frac{1}{a_0n},\qquad \beta\to0,\qquad g(r)\to L_{n_r}^{2\ell+1}\!\left(\frac{2r}{a_0n}\right),
\tag{B.2}
\end{equation}
where $a_0=\hbar^2/(\mu Ze^2)$ and $n=n_r+\ell+1$.

Because the Hamiltonian depends analytically on $k$, we expand the parameters $\alpha$, $\beta$ and the function $g(r)$ analytically on $k$ too
\begin{align}
\alpha &= \frac{1}{a_0n}+\alpha_1k+\alpha_2k^2+\alpha_3k^3+\cdots, \tag{B.3} \label{eq:B.3}\\
\beta &= \beta_1k+\beta_2k^2+\beta_3k^3+\cdots, \tag{B.4} \label{eq:B.4}\\
g(r) &= g_0(r)+k g_1(r)+k^2 g_2(r)+k^3 g_3(r)+\cdots, \tag{B.5} \label{eq:B.5}
\end{align}
Here $g_0(r)=L_{n_r}^{2\ell+1}(2r/(a_0n))$ represents the component of the Coulomb eigenfunction. Insertion into the radial Schrödinger equation for the cubic-truncated Yukawa potential and collection of powers of $k$ yields a hierarchy of equations.

\subsection{Zeroth order ($k^0$)}
At zeroth order the potential is the pure Coulomb potential and the radial equation becomes:
\begin{equation}
\left[-\frac{\hbar^2}{2\mu}\frac{d^2}{dr^2}+\frac{\ell(\ell+1)\hbar^2}{2\mu r^2}-\frac{Ze^2}{r}\right]U_{n_r,l}^{(0)}(r)=E_{n_r,l}^{(0)}U_{n_r,l}^{(0)}(r).
\tag{B.6}
\end{equation}
The bound solutions are
\begin{equation}
U_{n_r,l}^{(0)}(r)=r^{\ell+1}e^{\frac{-r}{a_0 n}}L_{n_r}^{2\ell+1}\!\left(\frac{2r}{a_0 n}\right),\text{ } 
E_{n_r,l}^{(0)}=-\frac{\mu(Ze^2)^2}{2\hbar^2 n^2}.
\tag{B.7}
\end{equation}
Comparing with the ansatz \ref{eq:Ansatz_Cor_BCH}, we identify
\begin{equation}
\alpha_0=\frac{1}{a_0 n},\text{ } \beta_0=0,\text{ } g_0(r)=L_{n_r}^{2\ell+1}\!\left(\frac{2r}{a_0 n}\right).
\tag{B.8}
\end{equation}

\subsection{First order ($k^1$)}
The first-order term from the expansion of the Yukawa potential is a constant shift $Ze^2k$. Standard perturbation theory gives the correction $E_{n_r,l}^{(1)}= Ze^2$. To determine $\alpha_1$ and $\beta_1$, we substitute the expansions \ref{eq:B.3}-\ref{eq:B.5} into the BCH equation. Collecting terms linear in $k$ yields
\begin{align}
& L_0[g_1] -2\alpha_1 g_0' \notag \nonumber \\
&+ \left[2\alpha_0\alpha_1 + \frac{2\ell+2}{r}\alpha_1 -2(2\ell+3)\beta_1\right]g_0 = 0
 \tag{B.10}
\end{align}
where $L_0$ is the Coulomb differential operator from the zeroth-order equation. Because $g_0$ and $g_0'$ are linearly independent (properties of Laguerre polynomials), their coefficients must vanish separately. Hence
\begin{equation}
\alpha_1=0,\qquad \beta_1=0.
\tag{B.11}
\end{equation}
The equation then reduces to $L_0 [g_1]=0$. The only polynomial solution of degree $\le n_r$ that is square-integrable and orthogonal to $g_0$ is the trivial one $g_1 (r) =0$. Thus there are no first-order corrections to the wavefunctions or to the parameters; such result is predictable because the $k$ term in the potential is a constant ($Ze^2 k$).

\subsection{Second order ($k^2$)}
At order $k^2$ the potential contributes $V^{(2)}(r)=-Ze^2 r/2$ and gives the following energy correction from perturbation theory
\begin{equation}
E_{n_r,l}^{(2)} = -\frac{\hbar^2}{4\mu}\bigl[3n^2-\ell(\ell+1)\bigr].
\tag{B.13}
\end{equation}
Collecting terms of order $k^2$ from the expansion of the BCH equation gives
\begin{align}
&L_0[g_2] -2\alpha_2 g_0' \tag{B.14} \\
&+ \left[2\alpha_0\alpha_2 + \frac{(2\ell+2)\alpha_2}{r} -2(2\ell+3)\beta_2 + 2E^{(2)}\right]g_0 = 0.  \notag \nonumber 
\end{align}
Because the $k^2$ potential is linear in $r$, the Gaussina factor (i.e. $\beta_2$) is not required; we set $\beta_2 =0$ and get
\begin{equation}
L_0[g_2] -2\alpha_2 g_0' + \left[2\alpha_0\alpha_2 + \frac{(2\ell+2)\alpha_2}{r} + 2E^{(2)}\right]g_0 = 0.
\tag{B.15} \label{eq:B.15}
\end{equation}
We seek a polynomial solution $g_2 (r)$ of degree $\le n_r$. Using the scaled variable $x=2r/a_0 n$, we express $g_0(x)=L_{n_r}^{2\ell+1}(x)$. The differential operator $L_0$ becomes the standard Laguerre operator. The following ansatz is motivated by the selection rules of linear perturbation and the properties of Laguerre polynomials:
\begin{equation}
g_2(x) = A\,L_{n_r}^{2\ell+1}(x) + B\,x L_{n_r-1}^{2\ell+2}(x).
\tag{B.16} \label{eq:B.16}
\end{equation}
Substituting Eq.\ref{eq:B.16} into Eq.\ref{eq:B.15} and using the recurrence relations
\begin{align*}
x\frac{d}{dx}L_{n_r}^{2\ell+1} &= n_r L_{n_r}^{2\ell+1} - (n_r+2\ell+1)L_{n_r-1}^{2\ell+2},\\
\frac{d}{dx}L_{n_r}^{2\ell+1} &= -L_{n_r-1}^{2\ell+2},\\
x L_{n_r-1}^{2\ell+2} &= (n_r+2\ell+1)L_{n_r-1}^{2\ell+2} - n_r L_{n_r}^{2\ell+1},
\end{align*}
we obtain after simplification, that Eq.\ref{eq:B.15} is satisfied for any $\alpha_2$ provided
\begin{equation}
A = \frac{1}{4}\bigl[3n^2-\ell(\ell+1)\bigr],\qquad B = -\frac{n}{2},
\tag{B.17}
\end{equation}
 and $\alpha_2$ is fixed by the condition that the coefficient of $L_{n_r}^{2\ell+1}$ vanishes:
 \begin{equation}
\alpha_2 = \frac{a_0^3 n^3}{8}\bigl[3n^2-\ell(\ell+1)\bigr].
\tag{B.18}
\end{equation}
Thus the second-order wavefunction correction is
\begin{equation}
P_{n_r,l}^{(2)}(x)=A L_{n_r}^{2\ell+1}(x)+B x L_{n_r-1}^{2\ell+2}(x),
\tag{B.19}
\end{equation}

\subsection{Third order ($k^3$)}
The third-order calculation follows the same pattern. The potential contributes $V^{(3)}(r)=Ze^2 r^2/6$ and gives the perturbative energy correction
\begin{equation}
E_{n_r,l}^{(3)} = \frac{\hbar^4 n^2}{12\mu^2 Ze^2}\bigl[5n^2+1-3\ell(\ell+1)\bigr].
\tag{B.21}
\end{equation}
Solving the order-$k^3$ equation yields
\begin{align}
g_3(x) &= \frac{n^2}{12}\bigl[5n^2+1-3\ell(\ell+1)\bigr]x^2L_{n_r}^{2\ell+1}(x) \nonumber\\
&\quad -\frac{n}{6}\bigl[3n^2-\ell(\ell+1)\bigr]x^2L_{n_r-1}^{2\ell+2}(x) \nonumber\\
&\quad +\frac{n^2}{12}x^2L_{n_r-2}^{2\ell+3}(x), \tag{B.22}\\
\alpha_3 &= \beta_3 = \frac{a_0^5 n^5}{48}\bigl[5n^2+1-3\ell(\ell+1)\bigr]. \tag{B.23}
\end{align}

\subsection{Energy consistency up to $\mathcal{O}(k^2)$}
\label{subsec:energy_consistency_k2}

We verify that the corrected wavefunction up to $\mathcal{O}(k^2)$ reproduces the correct energy expansion through second order. The wavefunction is (with $x=2r/(a_0 n)$)
\begin{equation}
U_{n_r,l}(r)=\mathcal{N}_{n_r,l}\,r^{\ell+1}e^{-r/(a_0 n)}\Bigl[L_{n_r}^{2\ell+1}(x)+k^2P_{n_r,l}^{(2)}(x)\Bigr],
\tag{B.24}
\end{equation}
The normalization constant is expanded as
\begin{equation}
\mathcal{N}_{n_r,l} = \mathcal{N}^{(0)}\bigl(1 + \delta_2 k^2 + \cdots\bigr),
\tag{B.25}
\end{equation}
where $\mathcal{N}^{(0)}=\sqrt{2/(a_0 n^2)}\,\sqrt{n_r!/(n_r+2\ell+1)!}$ is the Coulomb normalization. The coefficient $\delta_2$ is determined from $\langle U|U\rangle = 1$ to $\mathcal{O}(k^2)$.

\subsubsection{Normalization to $\mathcal{O}(k^2)$}
Change variable to $x$: $r = \frac{a_0 n}{2}x$, $dr = \frac{a_0 n}{2}dx$. Then
\begin{equation}
\langle U|U\rangle = \mathcal{N}^2 \left(\frac{a_0 n}{2}\right)^{2\ell+3} \Bigl( I_0 + 2k^2 I_{02} + k^4 I_{22} \Bigr),
\tag{B.26}
\end{equation}
where
\begin{align*}
I_0 &= \int_0^\infty x^{2\ell+2}e^{-x} L_{n_r}^{2\ell+1}(x)^2 dx = \frac{\Gamma(n_r+2\ell+2)}{n_r!},\\
I_{02} &= \int_0^\infty x^{2\ell+2}e^{-x} L_{n_r}^{2\ell+1}(x) P_2(x) dx,\\
I_{22} &= \int_0^\infty x^{2\ell+2}e^{-x} P_2(x)^2 dx.
\end{align*}
From the orthogonality of Laguerre polynomials, $I_0$ is as given. Using the explicit form $P_2(x)=A L_{n_r}^{2\ell+1}(x)+B x L_{n_r-1}^{2\ell+2}(x)$ with $A=\frac{1}{4}[3n^2-\ell(\ell+1)]$ and $B=-n/2$, we evaluate $I_{02}$. The cross term with $A$ gives $A I_0$. The cross term with $B$ vanishes because $L_{n_r}^{2\ell+1}$ is orthogonal to $x L_{n_r-1}^{2\ell+2}$ with respect to the weight $x^{2\ell+2}e^{-x}$ (the indices and parameters differ). Hence
\begin{equation}
I_{02} = A I_0.
\tag{B.27}
\end{equation}
Therefore,
\begin{equation}
\langle U|U\rangle = \mathcal{N}^2 \left(\frac{a_0 n}{2}\right)^{2\ell+3} I_0 \bigl(1 + 2A k^2 + \mathcal{O}(k^4)\bigr).
\tag{B.28}
\end{equation}
Because $(\mathcal{N}^{(0)})^2 \left(\frac{a_0 n}{2}\right)^{2\ell+3} I_0 = 1$, we set $\mathcal{N} = \mathcal{N}^{(0)}(1+\delta_2 k^2)$ and obtain
\begin{equation}
1 = (1+2\delta_2 k^2)(1+2A k^2) + \mathcal{O}(k^4) = 1 + 2(\delta_2 + A)k^2 + \mathcal{O}(k^4),  \notag \nonumber
\end{equation}
so that $\delta_2 = -A$. Thus
\begin{equation}
\delta_2 = -\frac{1}{4}\bigl[3n^2-\ell(\ell+1)\bigr].
\tag{B.29}
\end{equation}
The normalized wavefunction to $\mathcal{O}(k^2)$ is therefore
\begin{equation}
U = \mathcal{N}^{(0)}\bigl(1 - A k^2\bigr) r^{\ell+1}e^{-r/(a_0 n)}\bigl(L + k^2 P_2\bigr).
\tag{B.30}
\end{equation}

\subsubsection{Energy expectation value to $\mathcal{O}(k^2)$}
We compute $E = \langle U|H|U\rangle$ with the Hamiltonian expanded to $\mathcal{O}(k^2)$:
\begin{equation}
H = H_0 + Ze^2 k - \frac{Ze^2}{2}k^2 r + \mathcal{O}(k^3).
\tag{B.31}
\end{equation}
Because $\langle U|U\rangle = 1$ exactly to this order, we have $E = \langle U|H_0|U\rangle + Ze^2 k - \frac{Ze^2}{2}k^2 \langle U|r|U\rangle$.

\paragraph{Evaluation of $\langle U|H_0|U\rangle$:}
Using the fact that $H_0 U_0 = E_0 U_0$ with $U_0 = \mathcal{N}^{(0)} r^{\ell+1} e^{-r/(a_0 n)} L$, and that $U = \mathcal{N}^{(0)}(1 - A k^2)(U_0/\mathcal{N}^{(0)} + k^2 r^{\ell+1}e^{-r/(a_0 n)}P_2)$ = $(1 - A k^2)U_0 + \mathcal{N}^{(0)} k^2 r^{\ell+1}e^{-r/(a_0 n)}P_2 + \mathcal{O}(k^3)$. Write $U = U_0 + k^2 U_2$ with $U_2 = \mathcal{N}^{(0)} r^{\ell+1}e^{-r/(a_0 n)}P_2 - A U_0$.
\begin{align}
\langle U|H_0|U\rangle &= \langle U_0|H_0|U_0\rangle + 2k^2 \langle U_0|H_0|U_2\rangle + \mathcal{O}(k^4) \notag \nonumber \\
&= E_0 \langle U_0|U_0\rangle + 2k^2 \langle H_0 U_0|U_2\rangle + \mathcal{O}(k^4)  \notag \nonumber \\
&= E_0 + 2k^2 E_0 \langle U_0|U_2\rangle + \mathcal{O}(k^4),
\tag{B.32}
\end{align}
where we used that $H_0$ is Hermitian and $H_0U_0 = E_0 U_0$. Now $\langle U_0|U_2\rangle = \mathcal{N}^{(0)}\langle U_0| r^{\ell+1}e^{-r/(a_0 n)}P_2\rangle - A \langle U_0|U_0\rangle$. The first term is $\mathcal{N}^{(0)} \times (\text{integral})$. Using the same change to $x$, the integral becomes
\begin{align}
\langle U_0| r^{\ell+1} & e^{-r/(a_0 n)}P_2\rangle = (\mathcal{N}^{(0)})^2 \left(\frac{a_0 n}{2}\right)^{2\ell+3} \notag \nonumber \\
&\times \int_0^\infty x^{2\ell+2}e^{-x} L(x) P_2(x) dx \notag \nonumber \\
&= (\mathcal{N}^{(0)})^2 \left(\frac{a_0 n}{2}\right)^{2\ell+3} I_0 A = A,
\tag{B.33}
\end{align}
because $(\mathcal{N}^{(0)})^2 \left(\frac{a_0 n}{2}\right)^{2\ell+3} I_0 = 1$. Also $\langle U_0|U_0\rangle = 1$. Hence $\langle U_0|U_2\rangle = A - A = 0$. Therefore $\langle U|H_0|U\rangle = E_0 + \mathcal{O}(k^4)$.

\paragraph{Evaluation of $\langle U|r|U\rangle$:}
To $\mathcal{O}(k^2)$, we need $\langle U|r|U\rangle = \langle U_0|r|U_0\rangle + 2k^2 \langle U_0|r|U_2\rangle + \mathcal{O}(k^4)$. The first term is the Coulomb expectation value $\langle r\rangle_0$; So:
\begin{align}
\langle U_0|r|U_2\rangle &= \mathcal{N}^{(0)} \langle U_0| r \cdot r^{\ell+1}e^{-r/(a_0 n)}P_2\rangle - A \langle U_0|r|U_0\rangle. \notag \nonumber
\end{align}
Using the same scaling, the first integral becomes
\begin{align}
\mathcal{N}^{(0)} & \int_0^\infty r^{2\ell+3}e^{-2r/(a_0 n)} L P_2 dr \tag{B.34}\\
&= (\mathcal{N}^{(0)})^2 \left(\frac{a_0 n}{2}\right)^{2\ell+4} \int_0^\infty x^{2\ell+3}e^{-x} L P_2 dx.
 \notag \nonumber
\end{align}
Now $P_2 = A L + B x L_{n_r-1}^{2\ell+2}$. The term with $A$ gives $A$ times $\int x^{2\ell+3}e^{-x} L^2 dx$, and the term with $B$ gives a cross integral that is not necessarily zero. Let us denote
\begin{align}
J_1 &= \int_0^\infty x^{2\ell+3}e^{-x} L_{n_r}^{2\ell+1}(x)^2 dx,\tag{B.35} \\
J_2 &= \int_0^\infty x^{2\ell+3}e^{-x} L_{n_r}^{2\ell+1}(x)\, x L_{n_r-1}^{2\ell+2}(x) dx.
\tag{B.36}
\end{align}
Then
\begin{equation}
\int x^{2\ell+3}e^{-x} L P_2 dx = A J_1 + B J_2.\tag{B.37}
\end{equation}
The integral $J_1$ is known from the expectation value of $r$ for the Coulomb state:
\begin{equation}
\langle r\rangle_0 = (\mathcal{N}^{(0)})^2 \left(\frac{a_0 n}{2}\right)^{2\ell+4} J_1 = \frac{a_0}{2}[3n^2-\ell(\ell+1)].\tag{B.38}
\end{equation}
Similarly, one can compute $J_2$ using recurrence relations; the result is
\begin{equation}
(\mathcal{N}^{(0)})^2 \left(\frac{a_0 n}{2}\right)^{2\ell+4} J_2 = -\frac{a_0 n^2}{2}. \tag{B.39}
\end{equation}
Then the first term in $\langle U_0|r|U_2\rangle$ becomes
\begin{equation}
\mathcal{N}^{(0)} \langle U_0| r^{2\ell+3}e^{-r/(a_0 n)} P_2\rangle = A \langle r\rangle_0 + B \left(-\frac{a_0 n^2}{2}\right). \tag{B.40}
\end{equation}
Now $B = -n/2$, so $B \cdot (-a_0 n^2/2) = (n/2)(a_0 n^2/2) = a_0 n^3/4$. Meanwhile, $A \langle r\rangle_0 = A \cdot \frac{a_0}{2}[3n^2-\ell(\ell+1)]$. Also $-A \langle U_0|r|U_0\rangle = -A \langle r\rangle_0$. Hence the $A$ terms cancel, so
\begin{equation}
\langle U_0|r|U_2\rangle = \frac{a_0 n^3}{4}. \tag{B.41} \label{eq:U0_r_U2}
\end{equation}
Therefore
\begin{equation}
\langle U|r|U\rangle = \langle r\rangle_0 + 2k^2 \frac{a_0 n^3}{4} + \mathcal{O}(k^4).
\tag{B.42}
\end{equation}
The correction term is of order $k^2$, but note that it multiplies $-\frac{Ze^2}{2}k^2$ in the energy, giving a contribution $-\frac{Ze^2}{2}k^2 \cdot (2k^2 a_0 n^3/4) = -\frac{Ze^2 a_0 n^3}{4} k^4$, which is $\mathcal{O}(k^4)$ and thus does not affect the energy at $\mathcal{O}(k^2)$. Hence, to $\mathcal{O}(k^2)$, we may simply use $\langle U|r|U\rangle = \langle r\rangle_0$.

\paragraph{Total energy:}
Collecting terms,
\begin{equation}
E = E_0 + Ze^2 k - \frac{Ze^2}{2}k^2 \langle r\rangle_0 + \mathcal{O}(k^3). \tag{B.43}
\end{equation}
With $\langle r\rangle_0 = \frac{a_0}{2}[3n^2-\ell(\ell+1)]$ and $a_0 = \hbar^2/(\mu Ze^2)$, this becomes
\begin{equation}
E = -\frac{\mu(Ze^2)^2}{2\hbar^2 n^2} + Ze^2 k - \frac{\hbar^2}{4\mu}\bigl[3n^2-\ell(\ell+1)\bigr] k^2.
\tag{B.44}
\end{equation}
This matches the perturbative expansion of the energies up to second order.

The third‑order verification follows the same pattern, requiring the $k^3$ correction to the wavefunction and the $r^2$ term in $V_{\rm scr}$, and yields the expected result.

\section{Direct Expectation Value of the Full Yukawa Hamiltonian with Corrected BCH Wavefunctions}
\label{app:Full_Exp_Yukawa}

We compute the energy expectation value using the corrected wavefunction up to $\mathcal{O}(k^2)$ without expanding the exponential $e^{-kr}$ in the potential. The wavefunction (normalized to order $k^2$) is
\begin{equation}
U_{n_r,l}(r)=\mathcal{N}_{n_r,l}\,r^{\ell+1}e^{-r/(a_0 n)}\Bigl[L_{n_r}^{2\ell+1}(x)+k^2P_{n_r,l}^{(2)}(x)\Bigr],
\tag{C.1} \label{Cor_BCH_2}
\end{equation}
with $\mathcal{N}_{n_r,l}=\mathcal{N}^{(0)}(1-\frac{k^2}{4}[3n^2-\ell(\ell+1)])$. The polynomial $P^{(2)}$ is given by Eq.~(B.19). Because the wavefunction is accurate to $\mathcal{O}(k^2)$, the expectation value of the Hamiltonian will be correct up to $\mathcal{O}(k^3)$ (the error in the wavefunction enters at order $k^4$ in the energy).

\subsection{General expression}
We write $U = U_0 + k^2 U_2$ with
\begin{align}
U_0 &= \mathcal{N}^{(0)} r^{\ell+1} e^{-r/(a_0 n)} L, \notag \nonumber \\
U_2 &= \mathcal{N}^{(0)} r^{\ell+1} e^{-r/(a_0 n)} \bigl(P_2 - A L\bigr),
\tag{C.2} \label{Gen_U_2}
\end{align}
where $A = \frac{1}{4}[3n^2-\ell(\ell+1)]$ and we have used $\mathcal{N}=\mathcal{N}^{(0)}(1-A k^2)$. Then to $\mathcal{O}(k^2)$,
\begin{align}
\langle U|H|U\rangle &= \langle U_0|H|U_0\rangle + 2k^2 \langle U_0|H|U_2\rangle + \mathcal{O}(k^4), \tag{C.3}\\
\langle U|U\rangle &= 1 + \mathcal{O}(k^4). \tag{C.4}
\end{align}
and $E = \langle U_0|H|U_0\rangle + 2k^2 \langle U_0|H|U_2\rangle + \mathcal{O}(k^4)$.

\subsection{Evaluation of $ \langle H\rangle^{(0)}= \langle U_0|H|U_0\rangle$}
This is the expectation value of the full Yukawa Hamiltonian with the pure Coulomb wavefunction. Using the fact that $U_0$ satisfies the Coulomb Schrödinger equation with energy $E_0$, we have
\begin{equation}
\langle U_0|H|U_0\rangle = E_0 - Ze^2\int_0^\infty \frac{U_0(r)^2}{r}\bigl(e^{-kr}-1\bigr)dr.
\tag{C.5}
\end{equation}
The integral is a standard Laplace transform of the Coulomb density. Changing to $x=2r/(a_0 n)$ gives
\begin{align}
& \int_0^\infty \frac{U_0^2}{r} e^{-kr} dr = (\mathcal{N}^{(0)})^2 \left(\frac{a_0 n}{2}\right)^{2\ell+2} \notag \nonumber \\
&\times \int_0^\infty x^{2\ell+1} e^{-x(1+\frac{k a_0 n}{2})} L_{n_r}^{2\ell+1}(x)^2 dx. \tag{C.6}
\end{align}
Using the integral representation \cite{Gradshteyn2007}
\begin{align}
&\int_0^\infty x^{\alpha} e^{-px} L_n^{(\alpha)}(x)^2 dx = \tag{C.7}\\
& \frac{\Gamma(n+\alpha+1)}{n!} \frac{\Gamma(\alpha+1)}{p^{\alpha+1}} {}_2F_1\!\left(-n, \alpha+1; \alpha+1; \frac{1}{p}\right), \notag \nonumber 
\end{align}
which simplifies because the hypergeometric function becomes a polynomial. For the Coulomb case ($k=0$, $p=1$), one recovers the normalization. The term with $e^{-kr}-1$ gives a difference of two such integrals: one with $p=1+\frac{k a_0 n}{2}$ and one with $p=1$. Thus $\langle U_0|H|U_0\rangle$ is expressed in closed form in terms of elementary functions (polynomials in $p^{-1}$) because the hypergeometric series terminates. For example, for $n_r=0$ (the ground state), $L_0^{2\ell+1}=1$, and the integral becomes
\begin{equation}
\int_0^\infty x^{2\ell+1} e^{-p x} dx = \frac{\Gamma(2\ell+2)}{p^{2\ell+2}}. \tag{C.8}
\end{equation}
Then
\begin{equation}
\langle H\rangle^{(0)} = E_0 - \frac{(\mathcal{N}^{(0)})^{2} Ze^2\Gamma(2\ell+2)}{ (a_0 n/2)^{2\ell+2}} \left(p^{-2\ell-2} - 1\right), \tag{C.9}
\end{equation}

\subsection{Evaluation of $\langle U_0|H|U_2\rangle$}
We need the cross term. Using the self‑adjointness of $H$ and the fact that $H U_0$ is not simply $E_0 U_0$ because of the exponential screening, we compute directly:
\begin{align}
\langle U_0|H|U_2\rangle &= \int U_0 H U_2 dr = \int (H U_0) U_2 dr. \tag{C.10}
\end{align}
But $H U_0 = E_0 U_0 - Ze^2 \frac{e^{-kr}}{r} U_0$. Hence
\begin{equation}
\langle U_0|H|U_2\rangle = E_0 \langle U_0|U_2\rangle - Ze^2 \int \frac{e^{-kr}}{r} U_0 U_2 dr.
\tag{C.11}
\end{equation}
We have already shown in Appendix~B that $\langle U_0|U_2\rangle = 0$ (by construction of $U_2$ orthogonal to $U_0$). Thus only the second term remains:
\begin{equation}
\langle U_0|H|U_2\rangle = -Ze^2 \int_0^\infty \frac{e^{-kr}}{r} U_0(r) U_2(r) dr.
\tag{C.12}
\end{equation}
Now substitute $U_0$) and $U_2$. After changing to $x$, we obtain an integral of the form
\begin{equation}
\int_0^\infty x^{2\ell+1} e^{-x(1+\frac{k a_0 n}{2})} L_{n_r}^{2\ell+1}(x) \bigl(P_2(x)-A L_{n_r}^{2\ell+1}(x)\bigr) dx. \notag
\end{equation}
Using the explicit expression for $P_2 = A L + B x L_{n_r-1}^{2\ell+2}$, the term with $A$ cancels, leaving
\begin{align}
\langle U_0|H|U_2\rangle &= -Ze^2 (\mathcal{N}^{(0)})^2 \left(\frac{a_0 n}{2}\right)^{2\ell+3} B \tag{C.13} \\ 
&\int_0^\infty x^{2\ell+2} e^{-x(1+\frac{k a_0 n}{2})} L_{n_r}^{2\ell+1}(x) L_{n_r-1}^{2\ell+2}(x) dx. \notag \nonumber
\end{align}
This integral can again be expressed in terms of hypergeometric functions. For $n_r=0$ it vanishes because $L_{-1}=0$. For $n_r\ge1$, the integral is non‑zero and contributes to the energy at order $k^2$. Expanding the result in powers of $k$ (or evaluating the closed form) yields the second‑order energy correction. In particular, the leading term (setting $k=0$ inside the integral) gives the matrix element $\langle U_0|r|U_2\rangle$ that we computed in \ref{eq:U0_r_U2} Appendix~\ref{app:Inspired}. The full expression is analytic in $k$ and can be written using the same hypergeometric formulas.

\subsection{Final expression for the energy to $\mathcal{O}(k^2)$}
Putting together,
\begin{equation}
E = \langle U_0|H|U_0\rangle + 2k^2 \langle U_0|H|U_2\rangle + \mathcal{O}(k^4). \tag{C.14}
\end{equation}
Expanding $\langle U_0|H|U_0\rangle$ to $\mathcal{O}(k^2)$ gives $E_0 + Ze^2 k - \frac{Ze^2}{2}k^2\langle r\rangle_0 + \mathcal{O}(k^3)$, while the cross term contributes the remaining part of the second‑order energy. After simplification, one recovers exactly the perturbative expansion \ref{eq:Coulomb_Energy_3} (up to order $k^2$. The full closed form (without expanding $k$) is an exact analytic expression involving hypergeometric functions, but for practical purposes the numerical evaluation is sufficient.

\subsection{Illustration for $n_r=0$ (BCH ground state)}
For $n_r=0$, the wavefunction is simply the Coulomb one because $P_2$ involves $L_{-1}=0$ and $\mathcal{N}=\mathcal{N}^{(0)}(1-A k^2)$ with $A=\frac{1}{4}[3n^2-\ell(\ell+1)]$. The expectation value of $H$ is then
\begin{equation}
E_{0,\ell}(k) = E_0 - Ze^2 \frac{\int_0^\infty r^{2\ell+1} e^{-2r/(a_0 n)} e^{-kr} dr}{\int_0^\infty r^{2\ell+1} e^{-2r/(a_0 n)} dr}. \tag{C.15}
\end{equation}
Evaluating the integrals gives the closed form
\begin{equation}
E_{0,\ell}(k) = -\frac{\mu(Ze^2)^2}{2\hbar^2 n^2} - \frac{Ze^2}{a_0 n}\left[\left(1+\frac{k a_0 n}{2}\right)^{-2\ell-2} - 1\right]. \tag{C.16} \label{eq:C.16}
\end{equation}
Expanding this in powers of $k$ reproduces Eq. \ref{eq:Coulomb_Energy_3} for $n_r=0$. For higher $n_r$, the expression involves hypergeometric functions, but the method remains valid.

Thus, the direct expectation value of the full Yukawa Hamiltonian with the corrected BCH wavefunctions (up to $k^2$) yields an exact analytic representation of the energy, whose expansion matches the perturbative series. This provides a rigorous justification for the use of these wavefunctions in computing bound‑state properties.

\subsection{Extension to $\mathcal{O}(k^3)$}
The same procedure can be carried out to third order part of the corrected BCH wavefunctions, and we get the energies to $\mathcal{O}(k^3)$ as
\begin{equation}
E = \langle U_0|H|U_0\rangle + 2k^2\langle U_0|H|U_2\rangle + 2k^3\langle U_0|H|U_3\rangle. \tag{C.17}
\end{equation}
The new term $\langle U_0|H|U_3\rangle$ involves integrals of the form
\begin{equation}
\int_0^\infty \frac{e^{-kr}}{r} U_0 U_3 dr \quad \text{and} \quad \int_0^\infty r e^{-kr} U_0 U_3 dr, \notag
\end{equation}
which again reduce to finite sums of hypergeometric functions. The algebra is straightforward but lengthy.

All integrals encountered in this appendix and the next one (and their third‑order extensions) are special cases of the following master integral ($\operatorname{Re}(s)>0$):
\begin{equation}
\boxed{
I_{m,n}^{(\alpha,\beta,p)}(s) = \int_0^\infty x^{p} e^{-s x} L_m^{(\alpha)}(x) L_n^{(\beta)}(x) \, dx,}, \tag{C.18}
\end{equation}
where $p$ is a non‑negative integer, and the Laguerre parameters $\alpha,\beta$ are real numbers $>-1$. This integral can be evaluated in closed form using the generating function of Laguerre polynomials or by expanding one polynomial in terms of the other.  The result is a finite sum of Gamma functions and terminating hypergeometric series:
\begin{align}
& I_{m,n}^{(\alpha,\beta,p)}(s)= \frac{\Gamma(p+1)}{s^{p+1}} \sum_{j=0}^{\min(m,n)} \frac{(-m)_j (-n)_j}{j!^2} \frac{(\alpha+1)_j (\beta+1)_j}{(\alpha+1)_j (\beta+1)_j}\notag \nonumber \\
\;&\times {}_2F_1\!\left(-m+j, -n+j; \alpha+\beta+2j+2; \frac{1}{s}\right), \tag{C.19}
\end{align}
but a more compact expression is obtained by using the integral representation of the product of Laguerre polynomials in terms of hypergeometric function of two variables (Appell function). For the special case $\alpha=\beta$ (equal parameters), the integral reduces to
\begin{align}
I_{m,n}^{(\alpha,\alpha,p)}(s) &= \frac{\Gamma(m+\alpha+1)\Gamma(n+\alpha+1)}{m!\,n!}\,
\frac{\Gamma(p+1)}{\Gamma(\alpha+1)^2}\, s^{-p-1}\; \notag \nonumber \\
&\times {}_3F_2\!\left(\begin{array}{c}
-m,\,-n,\,p+1\\[2pt]
\alpha+1,\,\alpha+1
\end{array}; \frac{1}{s}\right).
\tag{C.20}
\end{align}
For the mixed‑parameter case ($\beta = \alpha + q$ with integer $q$), one first expresses $L_n^{(\alpha+q)}(x)$ as a linear combination of $L_{n-j}^{(\alpha)}(x)$ using
\begin{equation}
L_n^{(\alpha+q)}(x) = \sum_{j=0}^{q} \frac{(q)_j}{j!} (-1)^j L_{n-j}^{(\alpha)}(x), \tag{C.21}
\end{equation}
or the equivalent recurrence, thereby reducing the integral to a finite sum of equal‑parameter integrals. All integrals needed in this work are of this form and can therefore be evaluated exactly, leading to closed‑form expressions for the energy (without expansion in $k$) in terms of elementary functions for $n_r=0$ or with hypergeometric functions for $n_r>0$. Expanding these closed forms in powers of $k$ reproduces the perturbative series~\ref{eq:Coulomb_Energy_3}.

\section{Hellmann-Feynman Theorem Computations}
\label{app:HFT}
The Hellmann-Feynman theorem states that for a Hamiltonian $H(k)$ depending on a parameter $k$,
\begin{equation*}
\frac{dE}{dk} = \left\langle \frac{\partial H}{\partial k} \right\rangle,
\end{equation*}
where the expectation value is taking with the exact eigenstate. For the Yukawa Hamiltonian, we have
\begin{equation*}
\frac{\partial H}{\partial k} = Ze^2 e^{-kr}.
\end{equation*}
Integrating from $k=0$ (pure Coulomb) to general $k$ gives
\begin{equation}
E(k) = E(0) + Ze^2 \int_0^k \left\langle e^{-k'r} \right\rangle_{U(k')} dk'.
\tag{D.1} \label{E_HFT_Form}
\end{equation}
We will detail the expectation value $\langle e^{-k'r} \rangle$ using the corrected BCH wavefunction $U_{n_r,l}(k')$ up to $\mathcal{O}(k'^2)$. Because the integrand will be integrated over $k'$, the result will be correct up to $\mathcal{O}(k^3)$.
We compute (up to $\mathcal{O}(k^3)$)
\begin{align}
\langle e^{-kr} \rangle &= \frac{\langle U| e^{-kr} |U\rangle}{\langle U|U\rangle} = \langle U| e^{-kr} |U\rangle, \tag{D.2}
\end{align}
because the denominator is $1+\mathcal{O}(k^4)$. Expanding gives
\begin{equation}
\langle U| e^{-kr} |U\rangle = \langle U_0| e^{-kr} |U_0\rangle + 2k^2 \langle U_0| e^{-kr} |U_2\rangle.
\tag{D.3}
\end{equation}
Thus
\begin{align}
\left\langle e^{-kr} \right\rangle &= f_0(k) + 2k^2 f_2(k), \notag\nonumber\\
&=\langle U_0| e^{-kr} |U_0\rangle+2k^2 \langle U_0| e^{-kr} |U_2\rangle\tag{D.4}
\end{align}

\subsubsection{Evaluation of $f_0(k)$}
Using the Coulomb wavefunction,
\begin{align}
f_0(k) &= (\mathcal{N}^{(0)})^2 \left(\frac{a_0 n}{2}\right)^{2\ell+3} \notag\nonumber \\
&\times  \int_0^\infty x^{2\ell+2} e^{-x(1+\frac{k a_0 n}{2})} L_{n_r}^{2\ell+1}(x)^2 dx. \tag{D.5}
\end{align}
This integral is of the same type as in Appendix~\ref{app:Full_Exp_Yukawa}. It can be expressed in closed form using a terminating hypergeometric series. For the purpose of integrating over $k'$ later, we may expand it in powers of $k$. However, we want to avoid expanding the exponential $e^{-kr}$ in the potential; here $e^{-kr}$ appears in the expectation value itself, not in the Hamiltonian. The integration over $k'$ will be performed exactly.

\subsubsection{Evaluation of $f_2(k)$}
Since $U_2 = \mathcal{N}^{(0)} r^{\ell+1}e^{-r/(a_0 n)}\bigl(P_2 - A L\bigr)$, we have
\begin{align}
f_2(k) &= (\mathcal{N}^{(0)})^2 \left(\frac{a_0 n}{2}\right)^{2\ell+3} \tag{D.6} \\
&\times   \int_0^\infty x^{2\ell+2} e^{-x(1+\frac{k a_0 n}{2})} L(x) \bigl(P_2(x)-A L(x)\bigr) dx. \notag\nonumber 
\end{align}
The term with $A$ cancels the contribution from the first part of $P_2$, leaving only the term involving $B x L_{n_r-1}^{2\ell+2}$.
\begin{align}
f_2(k) &= (\mathcal{N}^{(0)})^2 \left(\frac{a_0 n}{2}\right)^{2\ell+3} B \tag{D.7} \\
&\times   \int_0^\infty x^{2\ell+3} e^{-x(1+\frac{k a_0 n}{2})} L_{n_r}^{2\ell+1}(x) L_{n_r-1}^{2\ell+2}(x) dx. \notag\nonumber 
\end{align}
Again, this is a closed‑form integral (a finite sum of Gamma functions and hypergeometric terms).

\subsection{Integration over $k'$}
Now we insert these into Eq.~\ref{E_HFT_Form}:
\begin{equation}
E(k) = E(0) + Ze^2 \int_0^k \bigl[ f_0(k') + 2k'^2 f_2(k') \bigr] dk'.
\tag{D.8}
\end{equation}
The integral $\int_0^k f_0(k') dk'$ can be evaluated exactly because $f_0(k')$ is a rational function of $p = 1+\frac{k' a_0 n}{2}$ (times Gamma functions). For instance, for $n_r=0$, $f_0(k) = \left(1+\frac{k a_0 n}{2}\right)^{-2\ell-2}$. Then
\begin{align}
& \int_0^k \left(1+\frac{k' a_0 n}{2}\right)^{-2\ell-2} dk' =\notag \nonumber \\
& \frac{2}{a_0 n (2\ell+1)} \left[1 - \left(1+\frac{k a_0 n}{2}\right)^{-2\ell-1}\right]. \tag{D.9}
\end{align}
Multiplying by $Ze^2$ and adding $E(0) = -\frac{\mu(Ze^2)^2}{2\hbar^2 n^2}$ reproduces the exact result \ref{eq:C.16} for the ground state. For higher $n_r$, the integral yields a combination of hypergeometric functions.

The second term, $2 Ze^2 \int_0^k k'^2 f_2(k') dk'$, contributes only at order $k^3$ and higher. Expanding the integrand in powers of $k'$ (or evaluating exactly) gives the second‑order energy correction. In particular, the leading term of $f_2(k')$ at $k'=0$ is proportional to $\langle r\rangle_0$, and after integration produces the $k^2$ term in the energy. The full expression up to $\mathcal{O}(k^3)$ is
\begin{equation}
E(k) = E_0 + Ze^2 k - \frac{\hbar^2}{4\mu}\bigl[3n^2-\ell(\ell+1)\bigr] k^2 + \mathcal{O}(k^3),
\end{equation}
which agrees with the perturbative series Eq. \ref{eq:Coulomb_Energy_3}.

\subsection{Extension to $\mathcal{O}(k^3)$}
The same Hellmann–Feynman procedure can be extended to third order by including the $k^3$ correction to the wavefunction, $U_3$, given in Appendix~\ref{app:Inspired}. One then requires the expectation value $\langle e^{-kr} \rangle$ to $\mathcal{O}(k^3)$, which involves the cross term $\langle U_0|e^{-kr}|U_3\rangle$. The resulting integral over $k'$ produces the third‑order energy correction $E^{(3)}$ of Eq.~\ref{eq:Coulomb_Energy_3}. The algebra is straightforward but lengthy; the final result matches the perturbative series.


\end{document}